\documentclass[pre,aps,twocolumn]{revtex4}

\usepackage{graphicx}
\usepackage{amsmath,empheq}
\usepackage{amssymb}
\usepackage{ifthen}
\usepackage{booktabs}
\usepackage{graphicx}
\usepackage{dcolumn}
\usepackage{bm}
\usepackage{longtable}
\usepackage{gensymb}

\newcommand{\bb}{\begin{equation}}
\newcommand{\ee}{\end{equation}}
\newcommand{\ba}{\begin{eqnarray*}}
\newcommand{\ea}{\end{eqnarray*}}
\newcommand{\rhor}{\rho({\bf r})}
\newcommand{\dd}{{\rm d}}
\newcommand{\rr}{{\mathbf r}}
\newcommand{\dr}{{\rm d}{\bf r}}

\begin{document}

\title{Capillary Condensation and Depinning Transitions in Open Slits}

\author{Alexandr \surname{Malijevsk\'y}}
\affiliation{ {Department of Physical Chemistry, University of Chemical Technology Prague, Praha 6, 166 28, Czech Republic;}
 {The Czech Academy of Sciences, Institute of Chemical Process Fundamentals,  Department of Molecular Modelling, 165 02 Prague, Czech Republic}}
 \author{Andrew O. \surname{Parry}}
\affiliation{Department of Mathematics, Imperial College London, London SW7 2BZ, UK}

\begin{abstract}
We study the low temperature phase equilibria of a fluid confined in an open capillary slit formed by two parallel walls separated by a distance $L$
which are in contact with a reservoir of gas. The top wall of the capillary is of finite length $H$ while the bottom wall is considered of
macroscopic extent. This system shows rich phase equilibria arising from the competition between two different types of capillary condensation,
corner filling and meniscus depinning transitions depending on the value of the aspect ratio $a=L/H$ and divides into three regimes: For long
capillaries, with $a<2/\pi$, the condensation is of type I involving menisci which are pinned at the top edges at the ends of the capillary. For
intermediate capillaries, with $2/\pi<a<1$, depending on the value of the contact angle the condensation may be of type I or of type II, in which the
menisci overspill into the reservoir and there is no pinning. For short capillaries, with $a>1$, condensation is always of type II. In all regimes,
capillary condensation is completely suppressed for sufficiently large contact angles which is determined explicitly. For long and intermediate
capillaries, we show that there is an additional continuous phase transition in the condensed liquid-like phase, associated with the depinning of
each meniscus as they round the upper open edges of the slit. Meniscus depinning is third-order for complete wetting and second-order for partial
wetting. Detailed scaling theories are developed for these transitions and phase boundaries which connect with the theories of wedge (corner) filling
and wetting encompassing interfacial fluctuation effects and the direct influence of intermolecular forces. We test several of our predictions using
a fully microscopic Density Functional Theory which allows us to study the two types of capillary condensation and its suppression at the molecular
level for different aspect ratios and contact angles.
\end{abstract}

\maketitle

\section{Introduction}


The statistical mechanics of inhomogeneous fluids, including the molecular theory of the interfacial region and the surface tension, the
formalisation and application of classical Density Functional Theory (DFT) and the study of novel surface phase transitions and critical phenomena
has received enormous interest in the last few decades \cite{rowlin, sullivan_gama, dietrich, schick, forgacs, hend, gelb, bonn}. Quite generally,
fluids are inhomogeneous whenever they are subject to an external potential, such as a phase separating gravitational field, or due to the
interaction with a confining solid substrate which is often modelled conveniently as an inert spectator phase, i.e., a wall or walls. Phenomena such
as wetting, including wetting transitions \cite{cahn, ebner, lipowsky83, lipowsky87}, capillary condensation  \cite{nakanishi81, nakanishi83,
evans84, evans85, evans86}, wedge filling \cite{hauge, rejmer, wood99, abraham02, delfino, binder03, bernardino, our_prl, our_wedge} and the thermal
Casimir effect \cite{fisher78, nightingale, kardar, herlein, sushkov, abraham13, paladugu} arise directly due to this confinement, and interactions,
with such walls. Studies of these phenomena are closely related to, and indeed partly grew out of, earlier work on finite-size effects at first and
second-order phase transitions, where for all but the imposition of periodic boundary conditions, surface effects are always present \cite{barber83,
binder83}. As well as being essential to the analysis of computer simulations of confined fluids, and Ising-like magnets, these studies have revealed
some very intriguing and deep properties associated with universal finite-size scaling near the bulk critical point \cite{cardy87, privman91}.

There is a third and more general context, however, where fluids must be modelled as being inhomogeneous, which is a marriage of the above two
scenarios. This occurs when we consider the interaction of a finite-size system which is open and in contact with a surrounding fluid reservoir. This
can be viewed as the study of edge effects since, in its simplest realisation, in addition to confining walls we must also allow for edges or corners
which demarcate the boundary between the micro or mesoscopic confined region with the macroscopic external environment. The purpose of the present
paper is to compare two of the simplest examples of this in which a fluid taken from a surrounding reservoir of vapour, at temperature $T$ and
chemical potential $\mu$ (or pressure $p$) can, at sufficiently low temperatures, condense between two parallel walls separated by a distance $L$. In
both cases the parallel walls produce an open capillary slit in contact with a bulk vapour. In one example, the walls are both of finite length $H$
and are perfectly adjacent (Fig.~1a). They are considered of macroscopic extent in all other directions. We refer to this as the $HH$ geometry which
has been studied recently \cite{fin_slit, geim20}. In the second scenario, however, which has not been studied previously, the bottom wall is also
considered to be of infinite extent representing perhaps a macroscopic table or work bench. We refer to this as the $H\infty$ geometry (Fig.~1b). In
both systems, at sufficiently low temperatures, the vapour between the walls will condense via a first-order phase transition to a liquid-like phase
as the pressure of the bulk gas is increased, happening before bulk saturation, $p_{\rm sat}$, is reached. This is the familiar phenomena of
capillary condensation. We remark at the outset that, beyond macroscopic and mean-field treatments, this first-order phase transition is rounded due
to finite-size effects although in practice it is of negligible importance provided we are away from the vicinity of the bulk critical temperature
(see later). We also remark that in our analysis we neglect the role played by gravity which is justified provided the separation $L$ of the walls is
much less than the capillary length parameter which, for molecular fluids, is of the order of mm. A preliminary account of some of our results has
appeared in \cite{prl}.


Our focus here centres on the location of the capillary condensation transition in these open systems and its dependence on the aspect ratio $a=L/H$
in each geometry. When the lateral extent of the slits is macroscopic, corresponding to $H=\infty$, or $a=0$, the location of the capillary
condensation in both geometries is the same and is very well described by the well-known Kelvin equation -- a macroscopic prediction which is known
to remain highly accurate even for microscopically narrow slits. When $H$ is finite however, the situation is more involved because of menisci which
appear near the open ends which separate the condensed liquid inside from the gas reservoir. In this case, for the $HH$ geometry it is known that the
location of the condensation transition is described, at least at a macroscopic level, by a generalised Kelvin equation which is characterised by an
edge contact angle $\theta_e$ \cite{fin_slit, fin_groove}. This describes the geometrical shape of the menisci which are always pinned at the
corners. The $H\infty$ geometry is, however, subtly different and leads to much richer behaviour.  The reason for this is that there are three
different components to the $H\infty$ geometry, each of which is associated with a phase transition. The slit brings with it the possibility of
capillary condensation, the two right-angle corners formed between the vertical sides and the (bottom) horizontal wall induce wedge filling and
finally associated with the two upper edges, at the open ends, is the possibility of meniscus depinning -- a new type of phase transition which we
describe here. Central to understanding the phase equilibria is the way in which the menisci at each end of the capillary connect with the bottom and
top walls. For example, it is clear they must meet the bottom (infinite) wall at the equilibrium Young contact angle $\theta$. The manner of the
connection with the upper wall however requires more attention, since the menisci may be pinned at each upper edge or unpinned, in which case the
liquid spills out of the capillary. In this case, the upper reaches of the menisci are in contact with the vertical sides of the walls -- a scenario
which must be present to connect with the phenomena of corner (wedge) filling. This additional phase transition must occur for sufficiently small
contact angles, $\theta<\pi/4$ as the pressure approaches bulk saturation. There are therefore two possible mechanisms for capillary condensation:
 \begin{itemize}
 \item type I in which the liquid remains inside the slit and the menisci are pinned at the upper edges,
 \item type II in which the liquid overspills into the reservoir and the two menisci are unpinned.
 \end{itemize}
We derive the generalised Kelvin equations for both these scenarios and show how they depend on the aspect ratio $a$. For type I condensation the
generalised Kelvin equation is determined by a new value of the edge contact angle $\theta_e$, distinct from that for the $HH$ geometry. The change
from type I to type II condensation is discussed in detail and expressed in terms of phase diagrams. Detailed connection with the scaling theory of
continuous wedge filling transitions is made. We also point out that for slits which exhibit type I condensation, as the pressure is increased
towards bulk saturation, the menisci eventually round the upper edges and therefore become unpinned. We show that at a macroscopic level such
meniscus depinning is a continuous phase transition which is third-order for complete wetting and second-order for partial wetting. Further, we
develop a scaling theory for its rounding at the mesoscopic level based on the theory of wetting transitions. Some of these predictions are verified
using a microscopic DFT model which allows us to view type I and type II condensations on the molecular level and show how capillary condensation is
suppressed for sufficiently large aspect ratios.

\begin{figure}
\includegraphics[width=8cm]{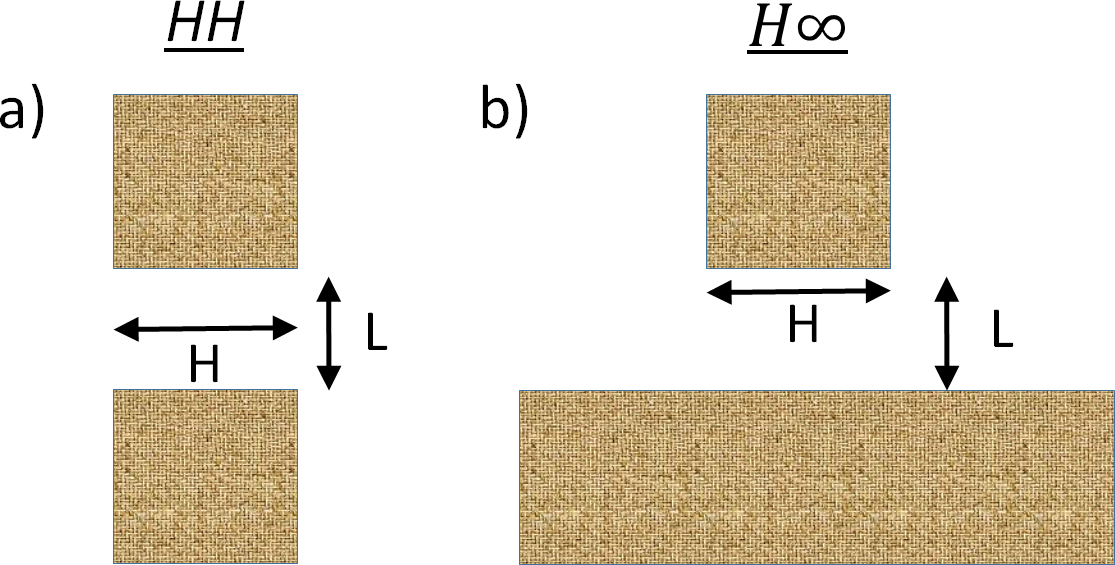}
\caption{Schematic illustration of two open finite-size slits formed by two parallel plates which are separated by a distance $L$ and which are in
contact with reservoir of gas. In the $HH$ geometry (a) both walls are of the same length $H$ so that the open ends have upper and lower edges. For
the $H\infty$ geometry (b) the lower wall is of infinite extent so there are just upper edges and the open ends of the capillary meet the reservoir
at two right angle corners. The dimensions of both capillaries are considered macroscopic in all other directions.} \label{sketch1}
\end{figure}

Our paper is arranged as follows. In section II, we begin with some introductory remarks about the modified Kelvin equation for the $HH$ geometry
before deriving the generalized Kelvin equations for the $H\infty$ system. In section III, we present these macroscopic results in terms of phase
diagrams and discuss the nature of meniscus depinning for complete and partial wetting. In section IV, we discuss the modifications occurring at the
mesoscopic level and develop detailed finite-size and cross-over scaling arguments which connect with the theories of wetting and wedge filling
transitions. In section V, we present the results of our microscopic DFT studies and end with a summary and discussion of future work.

\section{Kelvin equations for infinite and finite open slits}

In a narrow slit of width $L$, or cylindrical pore say, the forces of surface tension shift the phase boundary for coexistence between liquid and gas
away from the bulk saturation pressure $p_{\rm sat}$. The shift of the first-order phase boundary is very well described by the Kelvin equation,
which, although of macroscopic origin, is known to be accurate even for microscopically narrow slits. This is particularly true for partial wetting
where the Kelvin equation remains quantitatively accurate down to the molecular level. Let us begin by recalling the basic derivation and
interpretation of the Kelvin equation for a fluid confined between two identical infinite planar walls. Condensation occurs when the grand potential
$\Omega$ (per unit area of one of the walls, say) of the gas-like and liquid-like states are equal to each other. We assume here that the gas
condenses to liquid at a pressure $p_{\rm cc}$ below that of saturation, which will be the case if the contact angle $\theta$ is less than $\pi/2$.
For the gas-like state the volume and area contributions to the grand potential imply that for wide slits the grand potential is approximately given
by
\begin{equation}
\Omega_g= -p L +2\gamma_{\rm wg}\,,
\end{equation}
where $p$ is the pressure of a bulk reservoir of gas, assumed to be at chemical potential $\mu$ and temperature $T$, and $\gamma_{\rm wg}$ is the
wall-gas surface tension of a single wall. One approximation inherent here is that the surface tension term does not depend on the slit-width $L$
which is equivalent to neglecting the force of solvation between the walls. This is valid if we are away from the near vicinity of the capillary
critical point where the solvation force becomes long ranged. Similarly, for the liquid-like phase the grand potential can be written approximately
as
\begin{equation}
\Omega_l= -p^\dagger L +2\gamma_{\rm wl}\,,
\end{equation}
where $p^\dagger=p-\delta p$ is the pressure of the metastable bulk liquid and $\gamma_{\rm wl}$ is the wall-liquid surface tension. This again is
valid away from the capillary critical point and also provided we can ignore volume exclusion effects which arise when a high density liquid is
confined in a molecularly narrow slit. The difference between the grand potentials, $\Delta\Omega\equiv \Omega_g-\Omega_l$, is equal to
$\Delta\Omega=-\delta pL+2(\gamma_{wg}-\gamma_{wl})$. Setting $\Delta\Omega=0$ determines that the pressure shift at which condensation occurs is
$\delta p_{\rm cc}=2(\gamma_{\rm wg}-\gamma_{\rm wl})/L$ which leads to the famous Kelvin equation
\begin{equation}
\delta p_{\rm cc}=\frac{2\gamma\cos\theta}{L}\,,
\end{equation}
on using Young's equation $\gamma_{\rm wg}=\gamma_{\rm wl}+\gamma\cos\theta$ which defines the equilibrium contact angle of an infinite sessile drop.
Here, $\gamma$ is the liquid-gas surface tension. Hereafter, we express all our results using the convenient dimensionless reduced pressure shift
 \bb
\delta\tilde{p}\equiv\frac{L}{2\gamma}\delta p\,,
 \ee
  in terms of which the standard Kelvin equation simply reads
 \bb
  \delta \tilde{p}_{\rm cc}=\cos\theta\,.
 \ee
 Also, many of our results will be conveniently expressed in terms of the Laplace radius
  \bb
  R=\frac{\gamma}{\delta p}
  \ee
  of a circular meniscus.

As mentioned above, the Kelvin equation is particularly accurate for partial wetting. Corrections to it are present at the mesoscopic level for
complete wetting ($\theta=0$) where the singular contribution to the surface tension $\gamma_{\rm wg}$ arising from thick wetting films (or
equivalently the force of solvation between the liquid-gas interfaces and the walls) which reduce the effective slit width. A simple generalisation
of the Kelvin equation happens when the two walls are made of different materials, with distinct contact angles $\theta_1$ and $\theta_2$ say, in
which case the above argument leads to a generalised Kelvin equation
\begin{equation}
\delta \tilde{p}_{cc}=\frac{\cos\theta_1+\cos\theta_2}{2}\,. \label{costheta12}
\end{equation}
This will be a useful point of comparison for the geometry considered in the present paper even though the walls are materially identical. We note
that  the Kelvin equation has a simple geometrical interpretation since it identifies the unique pressure at which a circular meniscus of the Laplace
radius, which meets the walls at the appropriate equilibrium contact angle(s), phase separates the coexisting capillary-gas  and capillary-liquid
 phases.

\subsection{Capillary condensation in the $HH$ geometry}

\begin{figure}
\includegraphics[width=8cm]{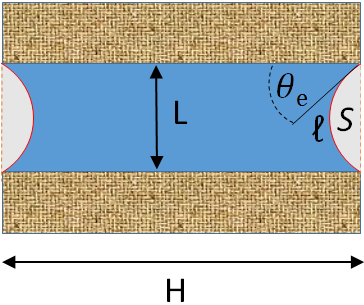}
\caption{Schematic illustration of a condensed capillary-liquid phase in the $HH$ geometry. Two circular menisci of Laplace radius $R=\gamma/\delta
p$ are pinned at the upper and lower edges which they meet at an edge contact angle $\theta_e$ which is pressure dependent and takes the value
$\theta_e^{\rm cc}$ at capillary condensation. The exposed area $S$ of gas and length $\ell$ of the menisci, needed in the determination of
$\theta_e^{\rm cc}$ for the generalised Kelvin equation $\delta\tilde p_{\rm cc}=\cos\theta_e^{\rm cc}$, are shown.}  \label{sketch_hh}
\end{figure}

We have recently extended these arguments in order to derive the generalised Kelvin equation for condensation in an open slit of width $L$ when the
(identical) walls are each of finite length $H$ \cite{fin_slit} (see Fig.~\ref{sketch1}a and Fig.~\ref{sketch_hh}). All other dimensions are
considered to be macroscopic and translational invariance is assumed in the direction normal to the cross-section in Fig.~\ref{sketch_hh}. For the
gas-like phase the contributions to the grand potential from the pressure and surface tension are similar to that for the infinite slit. Thus, per
unit length (into the wall), the grand potential can be well approximated by
\begin{equation}
\Omega_g= -pLH +2\gamma_{wg}H\,,
\end{equation}
where similar caveats about ignoring the force of solvation due to complete wetting layers or capillary criticality apply. There is no need to
consider the contribution from the outside (vertical) walls since these are identical for the gas and liquid-like states. For the liquid-like phase,
however, in addition to the (metastable bulk) pressure term and wall-liquid surface tensions, there is a new surface tension contribution from the
exposed area of two circular menisci. These {\it{must}} be present in an open pore and separate the capillary liquid from the gas reservoir. The
additional free-energy cost of these menisci increases the grand potential of the liquid-like phase implying that  for all finite $H$ the capillary
condensation must occur at a pressure which is {\it{closer}} to bulk saturation compared to that for the infinite slit. Recalling the geometrical
interpretation of the Kelvin equation, this means that the menisci cannot form a stable configuration within the slit and therefore must be pinned at
the open ends where each meets the corner at an edge contact angle $\theta_e$, distinct from $\theta$ and which will be pressure dependent. For the
liquid-like phase we can therefore write
\begin{equation}
 \Omega_l=-p^\dagger(LH-2S)+2H\gamma_{\rm wl}+2\gamma\ell\,,
\end{equation}
where
$\ell=(\pi-2\theta_e)R$ is the arc length of the menisci and $S=(\pi/2-\theta_e)R^2-\sin\theta_eRL/2$ is the area between the meniscus and the
open end. Elementary geometry implies that the radius $R$ of each meniscus must be related to the slit width by
 \bb
 \frac{L}{R}=2\cos\theta_e\,, \label{thetaeLR}
 \ee
which determines the value of $\theta_e$ for any pressure for which the liquid-like phase exists. The difference in the grand potentials of the
gas-like and liquid-like states is given by
 \bb
 \Delta\Omega=-\delta p(LH-2S)+2\gamma\cos\theta H-2\gamma\ell\,.
 \ee
 Setting $\Delta\Omega=0$ determines that capillary condensation in the $HH$ slit occurs when the
pressure shift is
\begin{equation}
\delta \tilde{p}_{\rm cc}(\theta, a)=\cos\theta_e^{\rm cc}\,,
\end{equation}
where $\theta_e^{\rm cc}$ is the value of the edge contact at condensation given by
\begin{equation}
 \cos\theta=\cos\theta_e^{\rm cc}+\frac{a}{2}\left[\sin\theta_e^{\rm cc}+\sec\theta_e^{\rm cc}\left(\frac{\pi}{2}-\theta_e^{\rm cc}\right)\right]\,,
 \label{costhetaeHH}
\end{equation}
where $a=L/H$ is the aspect ratio. Here, we have emphasized the dependence of the pressure shift on the contact angle $\theta$ and the aspect ratio
$a$.

For a finite length capillary, the value of the edge contact angle is always greater than the Young contact angle $\theta$ and approaches its value
only as $a\to0$. In general, this limit is approached analytically, 
 except for complete
wetting (see below). For long capillaries the pressure shift $\delta \tilde{p}_{\rm cc}$ can be written as an expansion in the aspect ratio,
\begin{equation}
\delta \tilde{p}_{\rm cc}(\theta, a)= \cos\theta -\frac{\alpha_1}{2} a -\frac{\alpha_2}{4} a^2+\cdots\,, \label{hh_exp}
\end{equation}
which highlights the corrections to the standard Kelvin equation. Here, the values of the coefficients are given by
\begin{equation}
\alpha_1=\left(\frac{\pi}{2}-\theta\right)\sec\theta+\sin\theta
\end{equation}
and
\begin{equation}
\alpha_2=\left(\frac{\pi}{2}-\theta\right)^2\sec^3\theta-\tan\theta\sin\theta\,.
\end{equation}
The higher-order terms in the expansion of $\delta \tilde{p}_{\rm cc}$ are analytic in the aspect ratio $a$ {\it{except}} for complete wetting which
reflects the non-analytic behaviour of the edge contact angle
\begin{equation}
\theta_e^{\rm cc}\approx\sqrt{\frac{\pi a}{2}} \label{cc_theta_e}
\end{equation}
for long slits. In this case the first three terms in the expansion of the pressure shift are
\begin{equation}
\delta \tilde{p}_{\rm cc}(0, a)=1 -\frac{\pi}{4}a -\frac{\pi^2}{16} a^2+\frac{\pi}{12}\sqrt{2\pi a^5}+\cdots\,, \label{exp_hh}
\end{equation}
which we shall return to. Quite generally, as the capillary is shortened, $\theta_e^{\rm cc}$ increases monotonically and reaches the value
$\theta_e^{\rm cc}=\pi/2$, at which $\delta \tilde{p}_{\rm cc}=0$, when the aspect ratio $a=a_0$ where
 \bb
 a_0=\cos\theta\,.\label{HH_a0}
 \ee
 For shorter capillaries the fluid inside the capillary {\it and} the gas reservoir simultaneously condense to liquid at bulk saturation $p_{\rm
sat}$, i.e., capillary condensation is suppressed for $a>a_0$ since the free-energy cost of creating the pinned menisci is too great.

\begin{figure}
\begin{center}
\includegraphics[width=8cm]{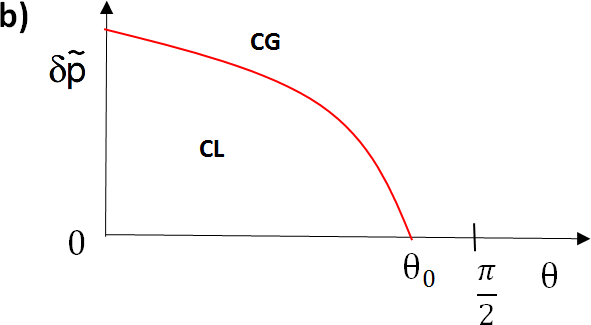}

\vspace*{0.3cm}

\includegraphics[width=8cm]{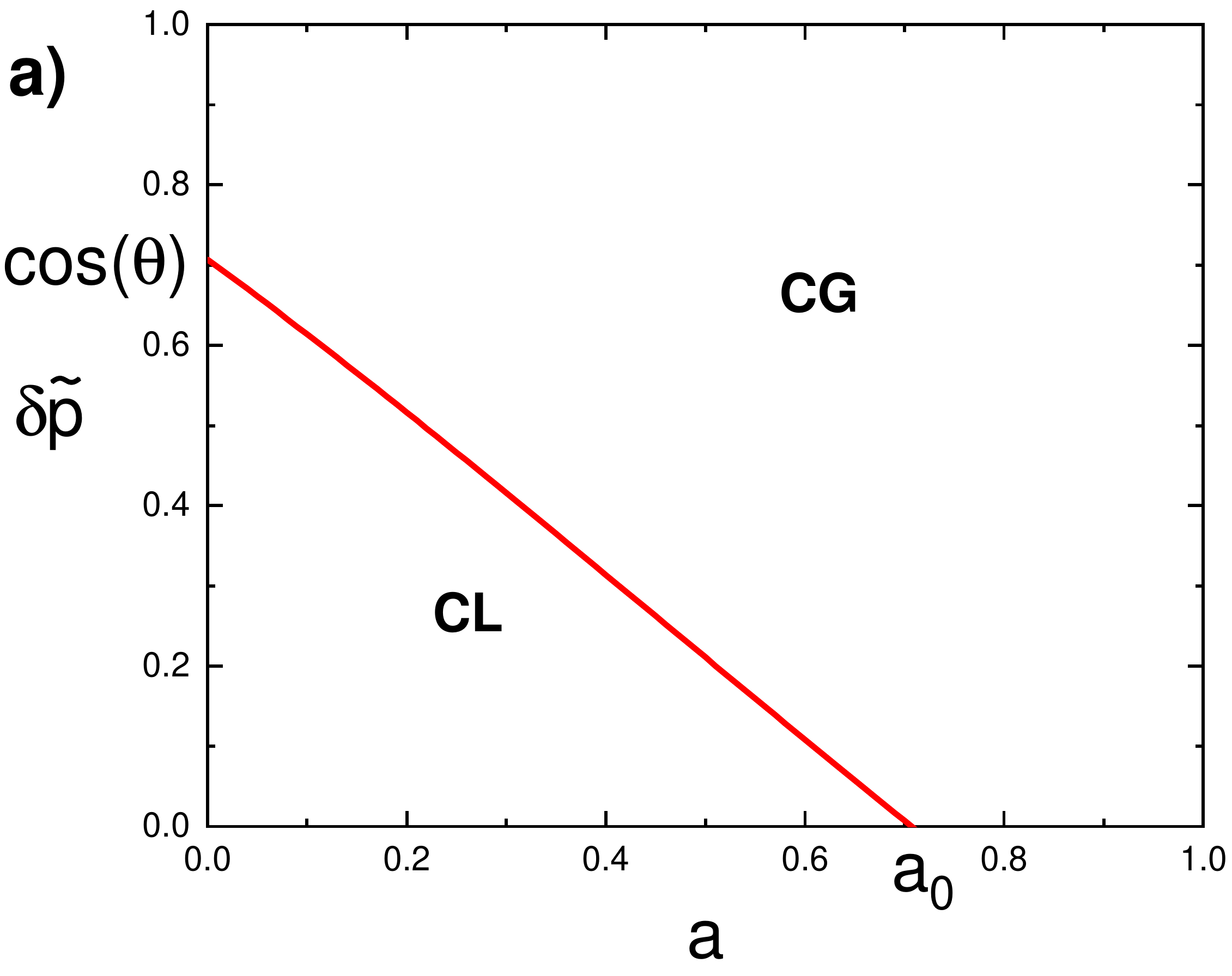}
\caption{Macroscopic phase diagrams for the $HH$ open capillary slit  showing the location of the capillary condensation line separating the stable
capillary-gas (CG) and capillary-liquid (CL) phases for a) different $\delta\tilde{p}$ and aspect ratios (for fixed contact angle $\theta$) and b)
different $\delta\tilde{p}$  and contact angle (for fixed aspect ratio). Capillary condensation is suppressed for sufficiently short capillaries with
aspect ratios $a>a_0=\cos\theta$ or, equivalently, for $\theta>\theta_0$ where $\theta_0=\cos^{-1}a$ in which case the fluid inside the capillary and
surrounding reservoir both condense to liquid at $p_{\rm sat}$. Capillary condensation is therefore suppressed in short capillaries when the aspect
ratio $a>1$.} \label{pd_hh}
\end{center}
\end{figure}

We can summarise these results in a simple phase diagram which shows the capillary condensation  phase boundary as a function of the aspect ratio
which separates the regions where the capillary gas (CG) and the capillary liquid (CL) phases are stable (see Fig.~\ref{pd_hh}a). Capillary
condensation only occurs for values of the aspect ratio up to $a_0$ beyond which it is suppressed. As $\theta$ is increased from zero so the line of
capillary condensation shrinks and vanishes at $\theta=\pi/2$ when the walls are neutral. We note that when $\theta>\pi/2$ the analogous phenomena of
capillary evaporation occurs when the bulk reservoir is liquid at pressure $p\ge p_{\rm sat}$. Alternatively, we can represent the capillary
condensation by plotting $\delta\tilde{p}$ vs $\theta$ for a given value of the aspect ratio (see Fig.~\ref{pd_hh}b). Here condensation is suppressed
for contact angles $\theta>\theta_0$ where, from (\ref{HH_a0}), it follows that $\cos\theta_0=a$. For long capillaries with $a\approx0$ the (red)
line of capillary condensation is described accurately by the expansion (\ref{hh_exp}). As the value of $a$ increases the whole line of capillary
condensation shrinks and vanishes as the aspect ratio is increased to unity since in this limit $\theta_0=0$. When the aspect ratio is close to
unity, the line of capillary condensation is described by
 \bb
 \delta \tilde{p}=1-a-\frac{\theta^2}{2}\,,
 \ee
 which ends at $\theta_0\approx \sqrt{2(1-a)}$.

\subsection{Capillary condensation in the $H\infty$ geometry}

\begin{figure}
\includegraphics[width=8cm]{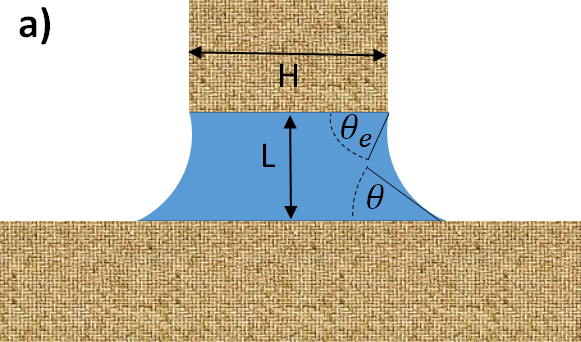}

\vspace*{1cm}

\includegraphics[width=8cm]{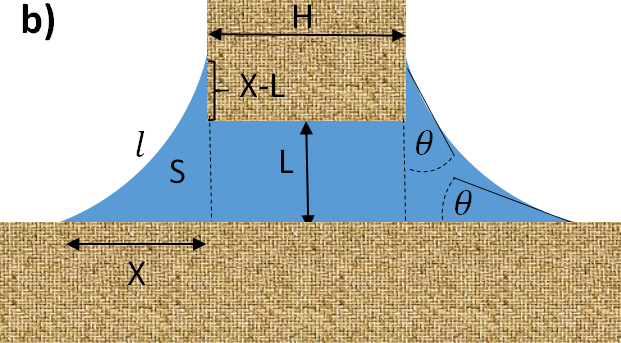}
\caption{Schematic illustration of two possible condensed capillary liquid phases in the $H\infty$ geometry. In the top panel (a) the two circular
menisci are pinned at the upper edges which they meet at an edge contact angle $\theta_e$ which is pressure dependent and takes the value
$\theta_e^{\rm cc}$ at type I capillary condensation. The bottom of the menisci meets the horizontal, lower wall at the equilibrium contact angle
$\theta$. In the lower panel (b), indicative of type II condensation, the two circular menisci are unpinned, spilling out into the right-angle
corners and meet the vertical and lower walls at the contact angle $\theta$.  In the lower panel we illustrate the meniscus length, $\ell$, overspill
area, $S$, and lateral extent along the bottom wall, $X$, and above the corner. Similar considerations apply when the meniscus is pinned. }
\label{sketch_hinf}
\end{figure}

With the above results for comparison, we now turn to the main subject of our paper which is the nature of condensation in an open slit in which one
of the walls (the bottom, say) is infinite while the other (the top, say) remains finite of length $H$ (see Fig.~\ref{sketch1}b). Again, we suppose
that the system is in contact with a bulk reservoir of gas at pressure $p$ (equivalently chemical potential $\mu$) at a temperature $T$ far below the
bulk critical point. We refer to this as the $H\infty$ geometry and will compare and contrast this to the $HH$ system described above. It is natural
to suppose once again that as the pressure is increased the fluid inside the capillary condenses to liquid at a pressure which is less than $p_{\rm
sat}$. Similar to the $HH$ geometry the equilibrium liquid-like phase is characterised by two circular menisci which separate the capillary liquid
from the bulk gas. For the same reasons as discussed earlier, at condensation itself (and indeed all higher pressures) these menisci cannot exist
within the slit and must be located near the open ends. The situation is, however, subtly different in several aspects which leads to richer phase
behaviour involving other interfacial phenomena. Since the bottom wall is infinite the menisci must meet it at Young's equilibrium contact angle
$\theta$. There are in principle, however, two possibilities for the upper part of each menisci. For sufficiently long slits the upper part of the
menisci connects with, and is pinned at, the edge, making an angle $\theta_e$ with the horizontal (upper) wall (see Fig.~\ref{sketch_hinf}a). This
new edge contact angle is pressure dependent for any CL phase but takes a specific value $\theta_e^{\rm cc}$ at capillary condensation (which we
stress is different to that defined for the $HH$ geometry). We refer to this as type I capillary condensation. For shorter capillaries, however {\it
and} for sufficiently small contact angles $\theta$ we shall show that the circular menisci are no longer pinned at the upper edges but rather sit
entirely outside the open ends and touch the bottom and vertical walls with the equilibrium contact angle $\theta$ (see Fig.~\ref{sketch_hinf}b). We
refer to this as type II capillary condensation.

\vspace*{0.5cm}

\subsubsection{Type I capillary condensation}

The location of type I capillary condensation may be determined, as for the $HH$ geometry, by comparing the grand potentials for the CG and CL
phases. We note that the edge contact angle of any CL phase is pressure dependent and is determined geometrically by
\begin{equation}
\frac{L}{R}=\cos\theta+\cos\theta_e\,, \label{LR}
\end{equation}
in contrast to the geometrical condition Eq.~(\ref{thetaeLR}) for the $HH$ geometry. The maximum value of the edge contact angle is
 \bb
\theta_e^{\rm max}=\theta+\pi/2\,,
 \ee
which is when the upper part of the menisci meet the vertical walls at Young's contact angle and are therefore unpinned. As for the $HH$ geometry,
the grand potential of the CL phase contains contributions arising from the volume of the metastable bulk liquid, the area of contact between the CL
and wall and finally the arc length of the menisci. These are readily calculated, and are slightly modified compared to the $HH$ geometry because the
menisci meet the corners and bottom walls at different angles and overspill out of the capillary. The difference in the grand potentials of the
gas-like and liquid-like phases is given by
 \bb
 \Delta\Omega=-\delta p(LH+2S)+2\gamma\cos\theta (H+X)-2\gamma\ell\,.
 \ee
Here the term $LH+2S$ is simply the total volume of liquid where
 \bb
 S= R^2 \left[\cos \theta \sin\theta_e -\frac{\sin 2\theta}{4} +\frac{\sin2\theta_e}{4} +\frac{1}{2}(\theta+\theta_e-\pi)\right]
 \ee
is the contribution from the overspill outside of each capillary end. Similarly, $2H+2X$ is the total area (per unit length) of contact between the
liquid and the walls where
 \bb
 X=R(\sin\theta_e -\sin\theta)
 \ee
 is the distance to which each meniscus extends along the bottom wall on both sides. Finally,
  \bb
 \ell=R(\pi-\theta-\theta_e)
  \ee
is the arc length of each meniscus involving both Young's contact angle and the edge contact angle (see Fig.~4). Setting $\Delta\Omega=0$ determines
that type I capillary condensation occurs when
\begin{equation}
\delta \tilde{p}_{\rm cc}^I(\theta, a)=\frac{\cos\theta+\cos\theta_e^{\rm cc}}{2}\,, \label{pcc_long}
\end{equation}
 where the value of the edge contact angle satisfies
\begin{equation}
\cos^2\theta=\cos^2\theta_e^{\rm cc} +a\frac{\pi-\theta-\theta_e^{\rm cc}+\sin(\theta+\theta_e^{\rm cc})}{1+a\tan\left(\frac{\theta_e^{\rm
cc}-\theta}{2}\right)}\,. \label{cosplus}
\end{equation}
Thus, the generalised Kelvin equation for type I condensation has a similar form to that for an infinite capillary with walls made of different
materials -- recall Eq.~(\ref{costheta12}). Simple inspection of (\ref{costhetaeHH}) and (\ref{cosplus}) shows that, for a given aspect ratio $a$,
there is no simple relation between the values of the edge contact angles  for the $HH$ and $H\infty$ geometries.
Given this, it is all the more surprising that when written as an expansion in the aspect ratio, the first two corrections to the standard Kelvin
equation for type I condensation in the $H\infty$ geometry are {\it{identical}} to that for the $HH$ geometry. That is,
\begin{equation}
\delta \tilde{p}_{\rm cc}^I(\theta, a)=\cos\theta -\frac{\beta_1}{2}a-\frac{\beta_2}{4}a^2+\cdots\,,
\end{equation}
where
\begin{equation}
\beta_1= \alpha_1\hspace{1cm}\beta_2=\alpha_2\,.
\end{equation}
Differences in the values of $\delta \tilde{p}_{\rm cc}$ for condensation in the $HH$ and $H\infty$  geometries only appear at the next-order in $a$
and are therefore near negligible for long capillaries. These are, again, non-analytic for complete wetting ($\theta=0$) where the upper edge contact
angle behaves as
\begin{equation}
\theta_e^{\rm cc}\approx\sqrt{\pi a}\,,
\end{equation}
as $a\to 0$, similar to Eq.~(\ref{cc_theta_e}) for the $HH$ geometry. Thus, for example, the expansion for the pressure shift for walls which are
completely wet is
\begin{equation}
\delta \tilde{p}_{\rm cc}^I(0, a)=1-\frac{\pi}{4}a -\frac{\pi^2}{16} a^2+\frac{\sqrt{\pi^3 a^5}}{6}+\cdots\,,
\end{equation}
which only differs from Eq.~(\ref{exp_hh}) in the coefficient of the $a^{5/2}$ term.

As the aspect ratio is increased, the loci of type I capillary condensation ends in one of two different ways depending on whether the contact angle
is greater or less than $\pi/4$ corresponding to the filling phase boundary for a right-angle corner \cite{hauge}. For $\theta<\pi/4$ type I
condensation ends when
\begin{equation}
 a_p=\frac{\cos(2\theta)}{\frac{\pi}{2}-2\theta}\,, \label{ap}
\end{equation}
which is the value of the aspect ratio at which the edge contact angle $\theta_e^{\rm cc}=\theta_e^{\rm max}$. At this point the menisci are no
longer pinned since they may be viewed as meeting the vertical walls at the equilibrium contact angle $\theta$. The corresponding value of the
pressure shift at this point is  $\delta \tilde{p}^I_{\rm cc}=(\cos\theta-\sin\theta)/2$. For large values of the aspect ratio the capillary
condensation is of type II, which does not involve any menisci pinning. For $\theta>\pi/4$ on the other hand the locus of type I condensation ends
when
 \bb
 a_0=\cot\theta\,, \label{a0}
 \ee
which is the value of the aspect ratio for which $\theta_e^{\rm cc}=\pi-\theta$, so that $\delta p_{\rm cc}^I=0$. We find it remarkable that the
capillary condensation at this terminus of type I condensation mimics the phase separation in an infinite slit where the walls are materially
different with opposing wetting properties, i.e., $\theta_2=\pi-\theta_1$ \cite{parry90}. For larger values of $a$ capillary condensation is
suppressed and the vapour within the walls and in the outside reservoir both condense to liquid at the same bulk phase boundary. This differs from
the value of $a_0$ defined for the $HH$ geometry.

We note that for fixed $a$ we can also define $\theta_p$ from solution of
 \bb
a=\frac{\cos(2\theta_p)}{\frac{\pi}{2}-2\theta_p}\,, \label{thetap}
 \ee
as the value of the contact angle at which type I condensation becomes type II. The value of $\theta_p$ is only defined for aspect ratios in the
range $2/\pi<a<1$ where the limiting values of the aspect ratios correspond to $\theta_p=0$ and $\theta_p=\pi/4$, respectively. Similarly, for aspect
ratios $a<1$ we can define a contact angle $\theta_0$ from
  \bb
  \theta_0=\cot^{-1}{a}\,,
  \ee
as the value of the contact angle at which type I condensation is suppressed. This will become clearer when we discuss the macroscopic phase
diagrams.

\subsubsection{Type II capillary condensation}

When $\theta<\pi/4$ and the aspect ratio is bigger than $a_p$ the capillary condensation takes on a different character. In this case the menisci are
no longer pinned at the corners and are each arcs  of circles which meet the bottom and vertical walls at the equilibrium Young contact angle
$\theta$. The difference between the grand potentials of the gas-like and liquid-like states is given by
\begin{equation}
\Delta\Omega=-\delta p(LH+2S)+2\gamma\cos\theta(H +2X -L)-2\gamma\ell\,.
\end{equation}
Here, as before, the term $LH +2S$ is the total volume of liquid which contains a contribution
\begin{equation}
S=  R^2 \left[\cos^2\theta -\frac{\sin 2\theta}{2} +\theta-\frac{\pi}{4}\right]
 \end{equation}
from the overspill at each end. Similarly, the term $2(H+2X-L)$ is the total contact area of the liquid with the wall, where
\begin{equation}
X=R(\cos\theta -\sin\theta)
\end{equation}
is the distance of the overspill along the bottom wall at each end. Note that the menisci also reach a distance $X-L$ above each corner which
contributes to the area of contact. Finally,
\begin{equation}
\ell= R\left(\frac{\pi}{2}-2\theta\right)
\end{equation}
is the arc length of each meniscus (see Fig.~4). None of these expressions involve an edge contact angle since there is no pinning. Setting $\Delta
\Omega=0$ determines that type II capillary condensation occurs at the pressure shift
\begin{equation}
\delta \tilde{p}_{\rm cc}^{II}(\theta, a)=\frac{a{\cal{A}}}{a-1+\sqrt{1+a^2-2a \left(\frac{\pi}{4}\sec^2\theta-\tan\theta\right)}} \label{pshort}
\end{equation}
    where the amplitude appearing in the numerator is
  \bb
    {\cal{A}}=\cos\theta-\sin\theta+\left(\theta-\frac{\pi}{4}\right)\sec\theta\,,
    \ee
which is simply $\cos\theta S/R^2$. When the aspect ratio $a=a_p$ this simplifies to $\delta \tilde{p}_{\rm cc}^{II}=(\cos\theta-\sin\theta)/2$ and
therefore provides continuity with the generalised equation Eq.~(\ref{pcc_long}) describing type I condensation for long capillaries. However, in
general, the expression (\ref{pshort}) does not have the form of a generalised Kelvin equation involving an effective contact angle that has an
obvious geometrical interpretation. We note that the amplitude ${\cal{A}}$ is only positive for $\theta<\pi/4$ which means that type II condensation
can never occur in the partial filling regime $\theta>\pi/4$.

The smallest value of the aspect ratio for which type II condensation occurs over the whole range of the contact angles (until it is suppressed for
$\theta>\pi/4$) is $a=1$, in which case the expression for $\delta \tilde{p}_{\rm cc}^{II}$ simplifies to
\begin{equation}
\delta \tilde{p}_{\rm cc}^{II}(\theta, 1)=\frac{{\cal{A}}}{\sqrt{2\left(1-\frac{\pi}{4}+\tan\theta-\frac{\pi}{4}\tan^2\theta\right)}}\,.
\label{pshort2}
\end{equation}

Some simplification also occurs when we suppose the wall is infinitesimally thin corresponding to the limit $H\to 0$ or equivalently $a\to\infty$. In
this case the location of type II capillary condensation has the limit
\begin{equation}
\delta \tilde{p}_{\rm cc}^{II}(\theta, \infty)=\frac{{\cal{A}}}{2}\,. \label{tilde_p_H0}
\end{equation}
Once again, this only gives a meaningful, positive, pressure shift for $\theta<\pi/4$ corresponding to the regime in which a right-angle corner is
completely filled at bulk saturation.

\section{Macroscopic phase diagrams}

\begin{figure*}
\includegraphics[width=8cm]{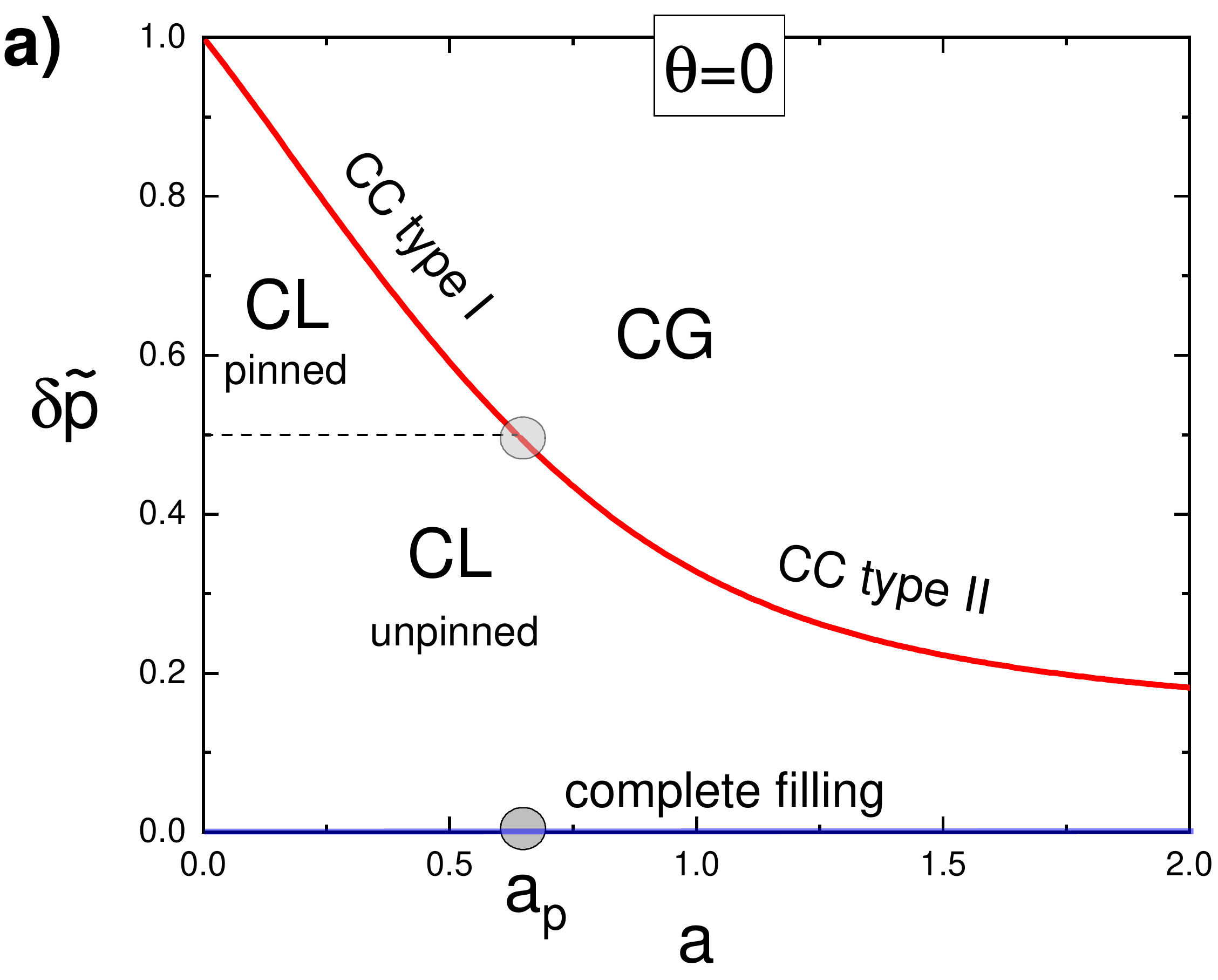} \hspace{0.5cm} \includegraphics[width=8cm]{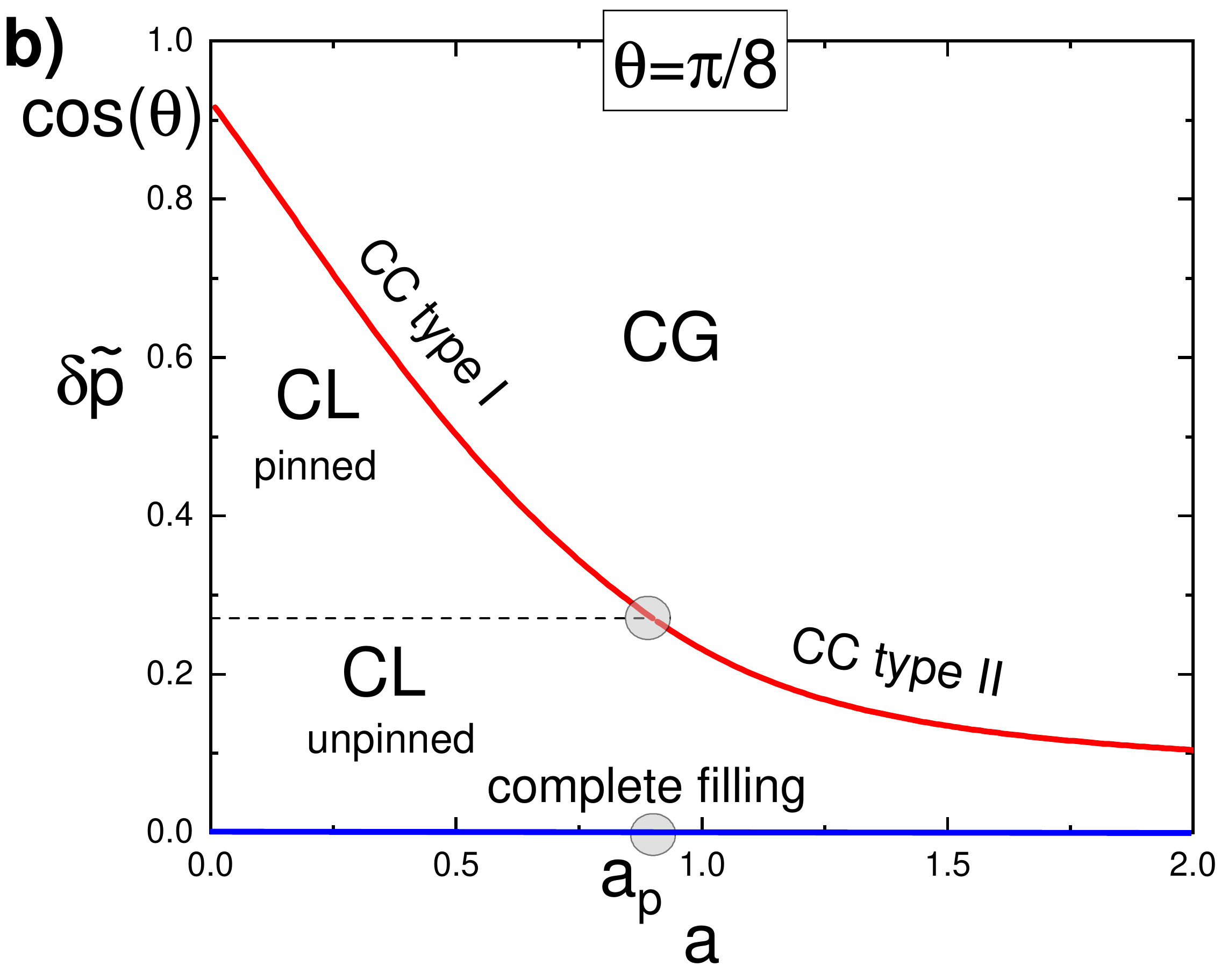}

\vspace*{0.5cm}

\includegraphics[width=8cm]{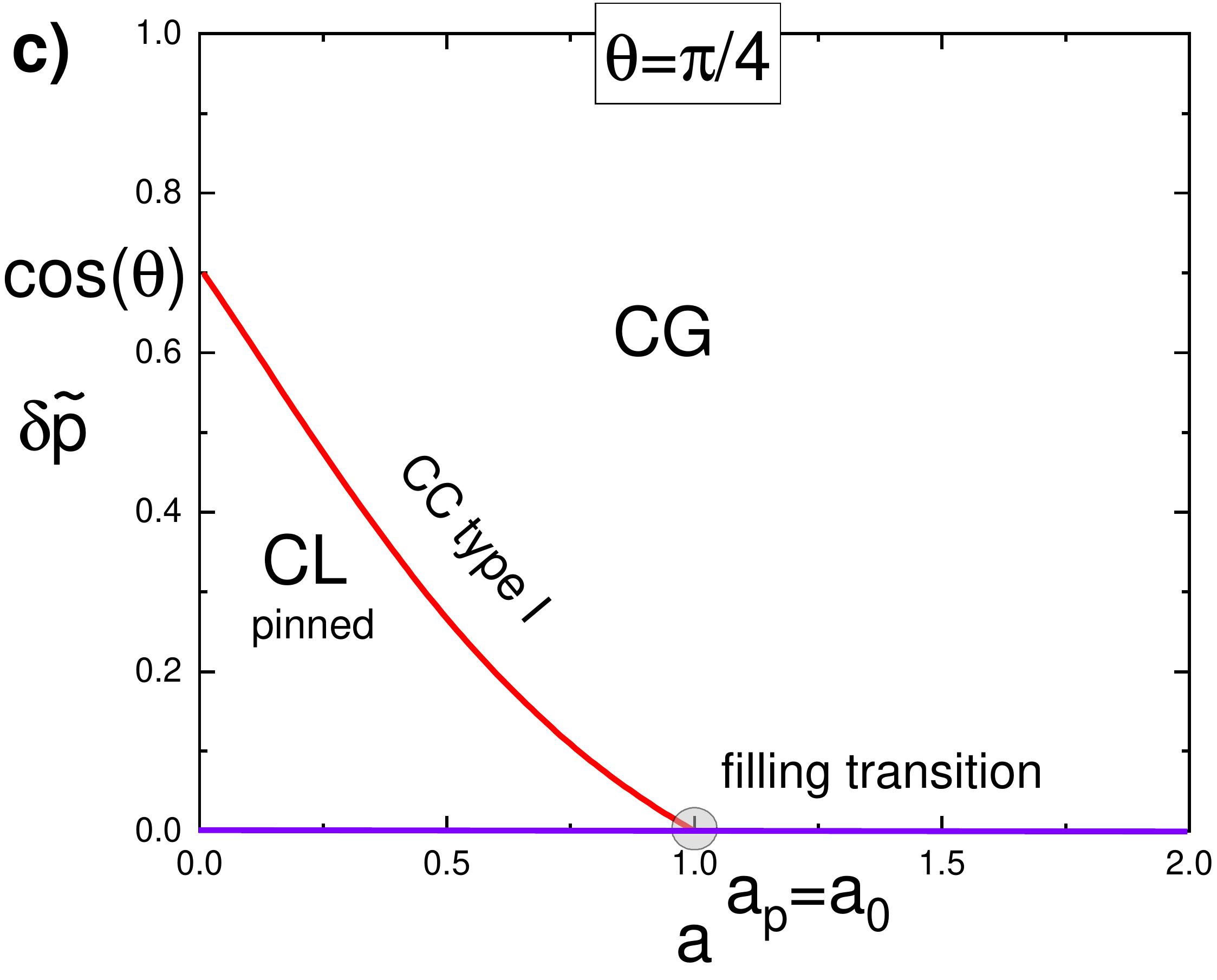} \hspace{0.5cm} \includegraphics[width=8cm]{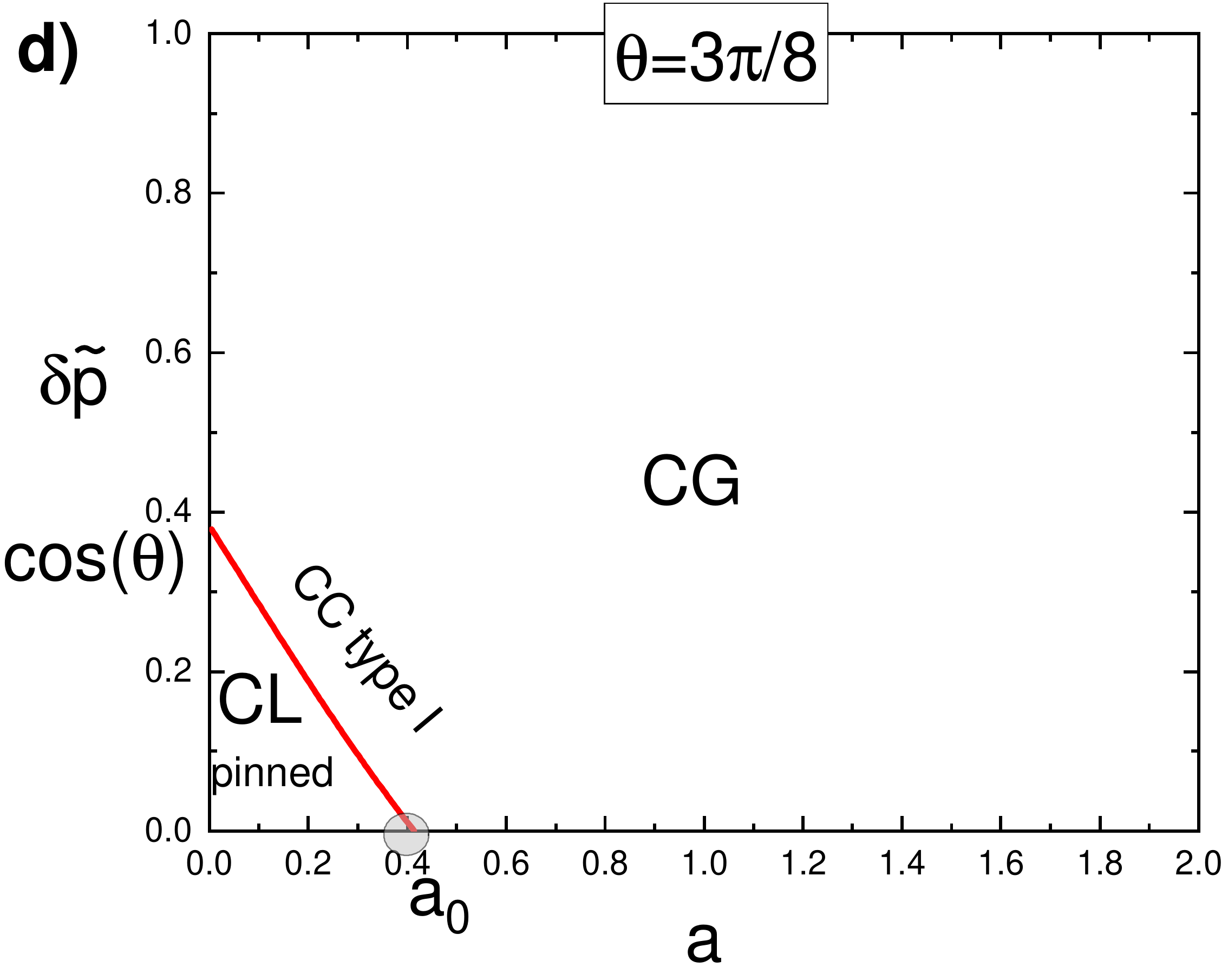}
\caption{Macroscopic phase diagrams showing the locations of type I and type II capillary condensation (CC) and the meniscus depinning transition
(dashed line) for the $H\infty$ geometry. Regions of stable capillary gas (CG), capillary-liquid (CL) as well as pinned and unpinned states are
shown. The four different phase diagrams represent a) $\theta=0$ complete wetting and complete corner filling. Here $a_p=2/\pi$ denotes where the
condensation changes character, b) $0<\theta<\pi/4$ corresponding to partial wetting but complete corner filling for which $2/\pi<a_p<1$ c) the
corner filling phase boundary $\theta=\pi/4$ at which the lines of meniscus depinning and type II condensation disappear by merging into the
saturation line $p=p_{\rm sat}$  and d) the partial corner filling regime, $\theta>\pi/4$ where only type I condensation occurs up to a maximum value
of the aspect ratio $a_0=\cot\theta$.} \label{pd_macro}
\end{figure*}

A convenient way of summarising the above macroscopic results is using the $\delta\tilde{p}$--$a$ and $\delta\tilde{p}$--$\theta$  phase diagrams, as
discussed earlier for the $HH$ geometry. These are now much richer reflecting the existence of different types of phase transitions and their phase
boundaries. We start by first considering the $\delta\tilde{p}$--$a$ phase diagram for fixed contact angle $\theta$.

\subsection{$\delta\tilde{p}$--$a$ phase diagrams}

We must consider two distinct ranges of the contact angle, $\theta<\pi/4$ and $\theta> \pi/4$ together with the marginal case $\theta=\pi/4$, which
corresponds to the corner filling phase boundary. We begin, however, with a discussion of the phase diagram for complete wetting, $\theta=0$, since
this is of particular physical importance and some of the expressions simplify.\\

{\indent{\bf Complete wetting $\boldsymbol{(\theta=0)}$}} (Fig.~\ref{pd_macro}a). There are two types of capillary condensation transition which
occur for $a<a_p$ and $a>a_p$ respectively where $ a_p$ takes its minimal value $a_p=2/\pi$. Again, these lines of capillary condensation (CC)
separate the regions where CG and CL  are the stable phases. For type I condensation the menisci are pinned at the top corners with an edge contact
angle $\theta_e^{\rm cc}<\pi/2$. The loci of type I condensation is described by the generalised Kelvin equation
\begin{equation}
\delta \tilde{p}_{\rm cc}^I(0, a)=\frac{1+\cos\theta_e^{\rm cc}}{2}
\end{equation}
where the value of the edge contact angle is found from solution of
\begin{equation}
\sin^2 \theta_e^{\rm cc}= a\,\left(\frac{\pi-\theta_e^{\rm cc}+\sin\theta_e^{\rm cc}}{1+a\tan\frac{\theta_e^{\rm cc}}{2}}\right)\,. \label{theta+0}
\end{equation}
 At $a=a_p=2/\pi$, the edge contact angle reaches its maximum value $\theta_e^{\rm cc}=\theta_e^{\rm max}=\pi/2$ and the menisci depin from the upper
edges. For $a>2/\pi$, type II condensation occurs and the menisci, at the transition, sit outside the capillary meeting the horizontal and vertical
walls tangentially in accordance with expectations of equilibrium complete wetting. The loci of type II condensation is described by
\begin{equation}
\delta \tilde{p}_{\rm cc}^{II}(0, a) = \frac{a(1-\frac{\pi}{4})}{a-1+\sqrt{1+a^2-\frac{\pi}{2}a}}\,,
\end{equation}
We note that these take the particular values $\delta \tilde{p}_{\rm cc}^{II}=1/2$ at $a=2/\pi$  and decreases monotonically to $\delta
\tilde{p}_{\rm cc}^{II}=(1-\pi/4)/2$ as $a\to\infty$.

We now return to the regime $a<a_p$ and consider the adsorption isotherm as the pressure is increased towards saturation.  For $\delta p>\delta
p_{\rm cc}^I$ the CG state is stable, while for  $\delta p<\delta p_{\rm cc}^I$ the CL state is stable. On increasing the pressure from the value at
$\delta p_{\rm cc}^I$ the edge contact angle  $\theta_e$ increases from $\theta_e^{\rm cc}$ according to the geometrical condition (\ref{LR})  until
it reaches its maximum allowed value $\theta_e=\theta_e^{\rm max}=\pi/2$, when the pressure shift takes the value
\begin{equation}
\delta \tilde{p}_{\rm md}=\frac{1}{2}\,, \label{p_tilde_md}
\end{equation}
at which point the meniscus depins (md). This is shown as the dashed line in Fig.~\ref{pd_macro}a and separates the regimes where the upper parts of
the meniscus are pinned (at the corners) or unpinned, in which case the menisci meet the vertical walls at some point above the corners. The pressure
shift (\ref{p_tilde_md}) is equivalent to the condition $R=L$, i.e. at the depinning transition the menisci each take the shape of a circle quarter
which just fit inside in the open ends of the capillary. The line of meniscus depinning meets the line of type I condensation at $a=a_p$. Crossing
the line of capillary condensation (be it type I or type II) corresponds to a first-order phase transition at which the adsorption jumps from to a
low to high value. Meniscus depinning on the other hand corresponds to a continuous phase transition where a higher derivative of the adsorption is
discontinuous. Let us consider this in detail. The adsorption (per unit length of the capillary) in the CL phase is simply the area of liquid
multiplied by $\Delta\rho$, the difference in the bulk liquid and gas densities. The adsorption of any CL phase always contains a trivial background
contribution $\Delta\rho H L$, arising from the area within the capillary, which we shall ignore, and an excess term arising from the two menisci
near the open ends. For the pinned CL phase, corresponding to $\delta p> \delta p_{\rm md}$ or equivalently $R<L$, the excess adsorption is
\begin{equation}
 \Gamma= 2 \Delta\rho R^2 \left[\sin \theta_e +\frac{\sin 2\theta_e}{4}+\frac{1}{2}(\theta_e-\pi)\right]\,, \label{Gammapincomplete}
 \end{equation}
where the edge contact angle $\theta_e$, for the present case of complete wetting, is given by
\begin{equation}
 \theta_e=\cos^{-1}\left(\frac{L-R}{R}\right)\,.
\end{equation}
For the unpinned CL phase, corresponding to $\delta p< \delta p_{\rm md}$ or equivalently $R>L$, on the other hand the excess adsorption is given
simply by
\begin{equation}
 \Gamma = 2\Delta\rho \left(1-\frac{\pi}{4}\right)R^2\,. \label{gamma_dep_trans}
 \end{equation}
Varying the pressure is equivalent to changing the radius of curvature $R$ and it is straightforward to check that the adsorption and its first
derivative, $\partial \Gamma/\partial R$  are continuous. However, the second derivative is discontinuous at $\delta \tilde{p}_{\rm md}$ with
 \bb
 \frac{\partial^2\Gamma}{\partial R^2}=\left\{\begin{array}{cc}
 \Delta\rho(4-\pi)\,;&\delta \tilde{p}=\delta \tilde{p}_{md}^-\,,\\
\Delta\rho(2-\pi)\,;&\delta \tilde{p}=\delta \tilde{p}_{md}^+\,.
\end{array}\right.
 \ee
Since the adsorption is proportional to the derivative of the grand potential $\Omega$ w.r.t. $\delta \tilde{p}$ we can anticipate that this
corresponds to a discontinuity in the third derivative of $\Omega$. This is indeed the case and a straightforward calculation determines that
 \bb
 \frac{\partial^3\Omega}{\partial R^3}=\left\{\begin{array}{cc}
 0\,;&\delta \tilde{p}=\delta \tilde{p}_{md}^-\,,\\
\frac{\gamma}{L^2}\,;&\delta \tilde{p}=\delta \tilde{p}_{md}^+\,.
\end{array}\right.
 \ee
 Thus, for complete wetting, $\theta=0$, meniscus depinning is a third-order phase transition.

Finally, as the pressure approaches $p_{\rm sat}$, the adsorption diverges due to the growth of two menisci each of which are quarter circles of
radius $R$. The horizontal blue line at $p=p_{\rm sat}$ shown in Fig.~\ref{pd_macro}a is therefore the line of complete corner filling. On
approaching this line the total adsorption diverges according to Eq.~(\ref{gamma_dep_trans}),
which is the universal, geometry determined, singularity for complete filling at a right angle corner \cite{rejmer, rascon2000}.\\

\begin{figure*}
\includegraphics[width=4.8cm]{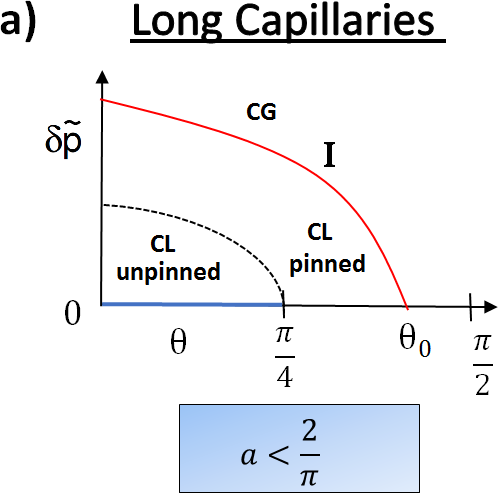} \hspace*{0.3cm} \includegraphics[width=5.5cm]{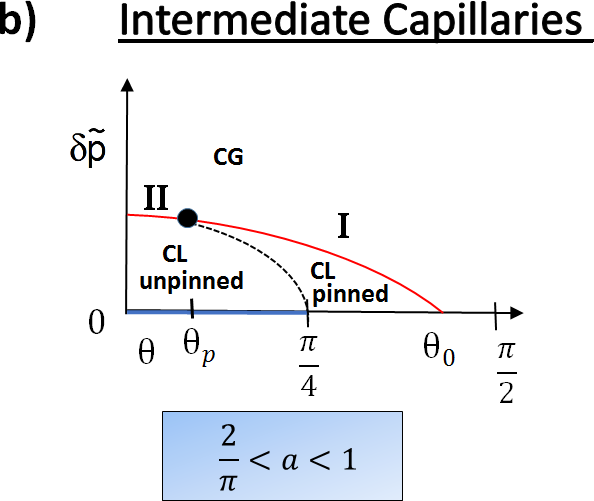} \hspace*{0.3cm}  \includegraphics[width=4.8cm]{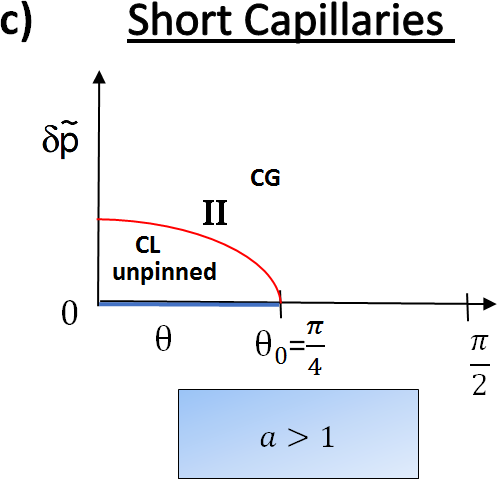}
\caption{Alternative illustration of the $H\infty$ macroscopic phase diagram $(\delta\tilde{p}, \theta)$ for the three relevant ranges of the aspect
ratio: a) {\bf long capillaries}, with $a<\frac{2}{\pi}$, the regime $a<2/\pi$ for which only type I condensation occurs up to a maximum value of the
contact angle $\theta_0=\cot^{-1}a>\pi/4$, beyond which capillary condensation is suppressed, b) {\bf intermediate capillaries}, corresponding to the
regime $2/\pi<a<1$ showing a change from type I to type II condensation at contact angle $\theta_p<\pi/4$ (determined from solution of
Eq.~(\ref{thetap})) and with condensation similarly suppressed for $\theta>\theta_0$ with $\theta_0>\pi/4$, and c) the {\bf short capillary} regime
$a>1$ for which only type II condensation occurs, which terminates at $\theta_0=\pi/4$.} \label{pd_theta-dp}
\end{figure*}

{\indent{\bf Complete filling $\boldsymbol{(0<\theta<\pi/4}$)} (Fig.~\ref{pd_macro}b). The phase diagram has the same qualitative structure for
$\pi/4>\theta>0$ as for complete wetting showing the two types of capillary condensation, the meniscus depinning, and the complete filling transition
at saturation. The loci of type I condensation is described by the generalised Kelvin equation (\ref{pcc_long}) with $\theta_e^{\rm cc}$ given by
(\ref{cosplus}) while type II is described by (\ref{pshort}). These meet at $a_p$ given by (\ref{ap}) which is now larger than $2/\pi$  which,
recall, is the value of the aspect ratio where $\theta_e^{\rm cc}=\theta+\pi/2$. Both lines of condensation lie closer to saturation than they do for
complete wetting as does the line of meniscus depining which occurs when
 \bb
 \delta \tilde{p}_{\rm md}=\frac{\cos\theta-\sin\theta}{2}\,, \label{cf_pmd}
 \ee
which again corresponds to a line of continuous phase transitions where $\theta_e=\theta_e^{\rm max}$. The location of this line does not depend on
the aspect ratio. Only in the region bounded by the loci of type I capillary condensation and the dashed meniscus depinning line  are the menisci of
the CL phase pinned at the top corners. It is intriguing to note, however, that for complete corner filling but partial wetting, the character of the
meniscus depinning transition is now different to that for complete wetting. For the pinned CL phase, corresponding to $\delta p> \delta p_{\rm md}$,
the adsorption is given by
\begin{equation}
\Gamma=2\Delta \rho R^ 2 \left[\cos \theta \sin\theta_e -\frac{\sin 2\theta}{4} +\frac{\sin2\theta_e}{4} +\frac{1}{2}(\theta+\theta_e-\pi)\right]\,,
\end{equation}
where $\theta_e$ is determined from (20). For the unpinned CL phase, on the other hand, for which $\delta p<\delta p_{\rm md}$, the adsorption is
given by
\begin{equation}
\Gamma=2\Delta \rho R^2 \left[\cos^2\theta -\frac{\sin 2\theta}{2} +\theta-\frac{\pi}{4}\right]\,. \label{cos2theta}
\end{equation}
It is straightforward to show that at the meniscus depinning transition ($\theta_e=\theta+\frac{\pi}{2}$) the adsorption is continuous. However, in
contrast to the earlier case of complete wetting, the first derivatives of the adsorption take different values at the phase boundary on the pinned
and unpinned sides. The difference between these is given by
\begin{equation}
\Delta \frac{\partial \Gamma}{\partial R}= 2\Delta\rho L\sin\theta(\tan\theta -1)\,.
\end{equation}
This means that for partial wetting the meniscus depinning transition is still continuous but of second-order. Note that when we set $\theta=0$, we
reproduce the result for complete wetting where the first derivative of $\Gamma$ is continuous.

Again, as $p\to p_{\rm sat}$, the circular menisci that sit outside the capillary grow in size and the total adsorption diverges according to
Eq.~(\ref{cos2theta}). 
This shows the same universal, geometry determined critical power law  $\Gamma\propto \delta p^{-2}$ as for complete wetting but with a smaller
amplitude ${\cal{A}}$. 

{\indent{\bf Filling phase boundary ($\boldsymbol{\theta=\pi/4)}$}} (Fig.~\ref{pd_macro}c). As the contact angle is increased towards the filling
phase boundary $\theta=\frac{\pi}{4}$ for a right-angle corner, the lines of meniscus depinning  and of type II capillary condensation collapse into
the saturation curve $p=p_{\rm sat}$ and the phase diagram takes a new qualitative form. Only type I capillary condensation, involving pinned
menisci, are possible and occur for aspect ratios up to a maximum value $a_0=a_p=1$. Capillary condensation is suppressed for all larger values of
$a$. What happens on approaching $p_{\rm sat}$, shown as the purple line, depends on the order of the filling transition and is not determined by the
present macroscopic considerations. To understand this we must turn to the theory of wedge filling and take into account the details of the
intermolecular forces. The simplest scenario is when the filling transition is first-order in which case at $p=p_{\rm sat}^-$ the menisci remained
pinned. These are now flat and simply connect the corners to the horizontal wall which they meet at angle $\theta=\frac{\pi}{4}$.

The most subtle case to consider is if the corner filling transition is second-order, which we leave to the next section when we discuss mesoscopic
effects for all the phase transitions described here.\\

{\indent{\bf Partial filling ($\boldsymbol{\theta>\pi/4)}$}} (Fig.~\ref{pd_macro}d).  When the contact angle $\theta>\pi/4$ the phase diagram is
simplest and is qualitatively the same as that for the $HH$ geometry (see Fig.~\ref{pd_hh}a). Only type I condensation exists  and occurs up to a
maximum value of the aspect ratio $a_0=\cot\theta$ (which as noted earlier is different to that defined for the $HH$ geometry). Capillary
condensation is suppressed for larger values of $a$. For $a<a_0$ the menisci of the capillary liquid phase flatten as $p$ is increased to $p_{\rm
sat}$. They remain pinned at the top edges, with an edge contact angle $\theta_e=\pi-\theta$, and meet the horizontal wall at the Young contact angle
$\theta$.

\subsection{$\delta\tilde{p}$--$\theta$ phase diagrams}

We may also represent the macroscopic predictions for the locations of type I and II capillary condensation using  $\tilde{p}$ vs $\theta$ projection
of the phase diagram for different values of the aspect ratio $a$ as shown in Fig.~\ref{pd_theta-dp}. The phase diagram is different depending on
whether the capillary is long, intermediate length or short. These are discussed separately:\\

{\bf Long capillaries} (Fig.~6a). If the aspect ratio $a<2/\pi$ then only type I condensation, involving pinned menisci, occur up to a maximum value
of the contact angle $\theta_0=\cot^{-1}a$ which lies in the range $\pi/4<\theta_0<\pi/2$. In addition, there is a line of continuous meniscus
depinning transitions, described by Eq.~(\ref{cf_pmd}), which ends at the corner filling phase boundary $\theta=\pi/4$. The saturation curve
$p=p_{\rm sat}$ itself breaks into complete corner filling ($\theta<\pi/4$), and partial corner filling ($\theta>\pi/4$) regions, for which the
adsorption diverges or remains finite, respectively.\\

{\bf Intermediate capillaries} (Fig.~6b). If the aspect ratio lies in the range $2/\pi<a<1$ the line of capillary condensation is of type II for
$\theta<\theta_p<\pi/4$, described analytically by Eq.~(\ref{pshort}), and type I for $\theta_p<\theta<\theta_0$ with $\theta_0>\pi/4$. The line of
meniscus depinning now only exists in the range $\theta_p<\theta<\pi/4$ since for smaller contact angles it occurs in a pressure regime for which the
CL phase is metastable. As the aspect ratio is increased to unity both $\theta_p$ and $\theta_0$ approach $\pi/4$ (from different sides) and the
lines of type I condensation and meniscus depinning vanish.\\

{\bf Short capillaries} (Fig.~6c). If the aspect ratio $a>1$ only type II capillary condensation, involving unpinned menisci, occurs up to a maximum
value $\theta_0=\pi/4$ beyond which capillary condensation is suppressed. We note that the short capillary regime begins when $a=1$ in which case the
line of type II condensation is described analytically by Eq.~(\ref{pshort2}).

\section{Mesoscopic considerations, rounding and scaling theory}

The above, purely macroscopic, considerations are due for some criticism. In particular, the capillary condensation, meniscus depinning transition
and type I/II crossover are rounded at the mesoscopic and molecular scale by thermal fluctuations and/or the direct influence of intermolecular
forces. The associated finite-size scaling of these transitions and phase boundaries is discussed in detail below:

\begin{enumerate}

\item {\bf Rounding of the Capillary Condensation Transitions}

Since the $HH$ and $H\infty$ geometries are pseudo one-dimensional then, strictly speaking, both type I and type II capillary condensation
transitions are rounded, due to thermal fluctuations, in accord with the well-developed theory of finite-size effects at first-order phase
transitions \cite{privman}. The transition from CG to CL is smooth, centered on $\delta p_{\rm cc}$ and rounded over a pressure range $\Delta p_{\rm
cc} \propto \exp(-\beta \gamma L H)$ where the factor $\gamma LH$ appearing in exponential is the approximate free energy cost of phase separating
the CG and CL along the capillary (normal to the cross-sections shown if Figs.~2 and 4). Such exponentially small rounding also applies to the
location of $a_0$ where capillary condensation is suppressed. At the pressure of condensation, the fluid breaks up into domains of CG and CL of
lengths of the order $\exp(\beta \gamma L H)$ along the capillary. Since the reduced surface tension is of order $\beta \gamma \sim 1/\xi_b^2$ where
$\xi_b$ is the bulk (liquid or gas) correlation length, away from the bulk critical temperature the rounding of the capillary condensation
transitions and location of $a_0$ in either the $HH$ or $H\infty$ geometries is negligible unless either dimension $L$ or $H$ is molecularly small.
Such rounding is only of significance in the near vicinity of the capillary (pseudo) critical temperature $T_c(L,H)$ which marks the end of the
pseudo phase coexistence. This capillary critical point itself will occur approximately when the smallest of the dimensions $L$ or $H$ is of order
$\xi_b$. We note that the rounding of the condensation transition and rounding of $a_0$ where condensation is suppressed are entirely absent in
mean-field DFT treatments of the phase equilibria where the phase transitions remain sharp.

\item {\bf Vanishing of type II Condensation near the Corner Filling Phase Boundary}

\begin{figure}
\includegraphics[width=6cm]{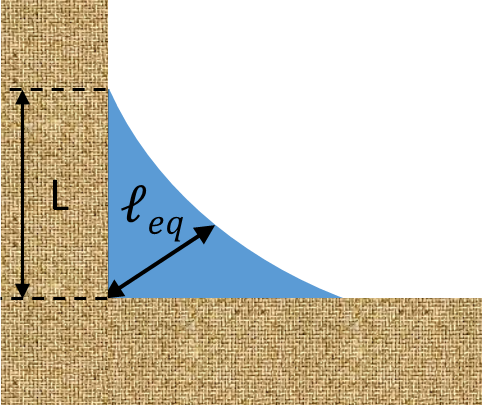}
\caption{Length-scales relevant for the finite-size scaling for the vanishing of type II condensation. Schematic illustration of the microscopic
adsorption of liquid at a right-angle corner, in the absence of the slit opening. At a continuous wedge/corner filling transition the thickness
$\ell_{\rm eq}$ of liquid diverges continuously $\ell_{\rm eq}\propto (\theta-\pi/4)^{-\frac{1}{2}}$ as the contact angle decreases to the filling
phase boundary at $p=p_{\rm sat}^-$. When the vertical extent $\sqrt{2}\ell_{\rm eq}$ of the adsorbed layer is larger than the slit width $L$, shown
as dashed lines, the macroscopic phase diagrams are modified, such that meniscus depinning and type II condensation persist into the partial filling
regime until $\theta-\pi/4\propto 1/L^2$. } \label{wedge}
\end{figure}

The macroscopic $\delta\tilde{p}$--$a$ phase diagrams for the $H\infty$ geometry change qualitatively precisely at the corner filling phase boundary
$\theta=\pi/4$. As discussed earlier, starting from the case of complete wetting, the lines of meniscus depinning and type II condensation both merge
into the saturation curve $p_{\rm sat}$ as the contact angle increases towards $\theta=\pi/4$. This macroscopic prediction remains valid provided the
smallest mesoscopic length associated with the corner filling transition is much smaller than the slit width $L$. Thus, for walls that exhibit
first-order corner filling transitions this prediction remains accurate down to the molecular scale since the adsorption of liquid at the right-angle
corners, subtended between the (upper) vertical sides and (bottom) horizontal wall, remains microscopically small at $p=p_{\rm sat}$ and
$\theta=\pi/4$. However, for walls that show continuous, second-order, corner filling transitions the cross-over from the $\theta<\pi/4$ phase
diagram to the $\theta>\pi/4$ phase diagram requires more careful consideration since even for contact angles close to but greater than $\pi/4$ the
microscopic adsorption of liquid at the right-angle corners will be significant. The crossover from the $\theta<\pi/4$ phase diagram to the
$\theta>\pi/4$ phase diagrams may be understood by appealing to the microscopic theory of corner filling transitions which we discuss here
specifically for the case of systems with dispersion forces. At a single right angle corner, see Fig.~\ref{wedge}, the microscopic thickness
$\ell_{\rm eq}$ of the adsorbed layer of liquid may be found from minimizing the appropriate corner contribution to the excess free-energy
\cite{rejmer}
 \begin{equation}
 F_{\rm corner}=\delta p \ell^2 +t \ell +\frac{A}{\ell} \label{Fcorn}
 \end{equation}
where all unimportant constants and proportionality factors have been ignored and $t=\theta-\pi/4$ is the temperature-like scaling variable for the
filling transition. The first term in this expression is the volume contribution arising from the metastability of the liquid when $p>p_{\rm sat}$,
the second term arises from all the surface tension contributions and changes sign at the thermodynamically determined filling phase boundary while
the last term is the direct influence of the dispersion forces with $A$ the Hamaker constant which is positive for continuous wedge filling.
Minimizing (\ref{Fcorn}) therefore determines the equilibrium value of the microscopic thickness of the adsorption of liquid at the right-angle
corner:
 \begin{equation}
 2\delta p\ell_{\rm eq} +\left(\theta-\frac{\pi}{4}\right) =\frac{A}{\ell_{\rm eq}^2}\,. \label{lcorn}
 \end{equation}
Provided this is smaller than $L$ then the macroscopic prediction for the phase diagram in the partial filling regime $\theta>\pi/4$ remain accurate.
However, the loci of both the meniscus depinning and type II condensation must follow from the finite-size scaling condition that $\sqrt{2}\ell_{\rm
eq}\approx L$ since both transitions must still be present when there is overspilling into the gas reservoir. It therefore follows that for walls
that show continuous corner filling the crossover from the complete to partial filling regimes is as follows:

\begin{itemize}

\item Exactly at the corner filling phase boundary $\theta=\pi/4$, similar to Fig.~5b, there is still a remnant of the meniscus depinning and type II
condensation which both lie along the lines $\delta p_{\rm md}\sim \delta p_{\rm cc} \propto A/L^3$. On approaching the purple line the corner
menisci grow continuously and the adsorption diverges as $\Gamma\propto \delta p^{-2/3}$ in accord with the predictions for the critical isotherm for
continuous corner filling. We can place this simple finite-size scaling argument in a more general setting using the scaling theory of continuous
wedge filling transitions \cite{parry2000}. In the vicinity of a wedge filling phase boundary, the thickness of the adsorbed layer of liquid shows
scaling behaviour $\ell_{\rm eq}\approx t^{-\beta_w} \Lambda_w (\delta p t^{-\Delta_w})$ where $\beta_w$ and $\Delta_w$ are the film thickness and
gap exponent respectively and $\Lambda_w(x)$ is a scaling function. Along the filling critical isotherm, equivalent to setting $\theta=\pi/4$ for a
right-angle corner, the film thickness therefore must diverge as $\ell_{\rm eq} \approx \delta p^{-\beta_w/\Delta_w}$ on approaching the pressure of
bulk saturation. Thus, we anticipate that in the $H\infty$ geometry the mesoscopic remnant of the type II capillary condensation and meniscus
depinning transitions occur when
 \bb
\delta p_{\rm md}\sim\delta p_{\rm cc}\propto L^{-\frac{\Delta_w}{\beta_w}}\,,
 \ee
which recovers the above results for dispersion forces on using the appropriate critical exponents $\beta_w=\frac{1}{2}$ and $\Delta_w=\frac{3}{2}$.
For systems with short-ranged forces this scaling argument predicts $\delta p_{\rm md}\sim \delta p_{\rm cc}\propto L^{-5}$ on substituting for the
universal, fluctuation-dominated, values of the critical exponents $\beta_w=\frac{1}{4}$ and $\Delta=\frac{5}{4}$.

\item In the small temperature window $0<t< A/L^2$ in the partial corner filling regime the lines of meniscus depinning type II condensation lie
along  $\delta p_{\rm md}\sim\delta p_{\rm cc}\propto A/L^3-t/L$. The adsorption of liquid at the corners remains microscopic as bulk saturation is
approached.  As the contact angle is increased the lines of meniscus depinning and type II condensation eventually merge with the saturation curve
recovering the macroscopic phase diagram Fig.~5c when $t\sim A/L^2$.

Using the same scaling argument described above we anticipate that, more generally, the meniscus depinning and type II capillary condensation
disappear by merging into the saturation curve $p=p_{\rm sat}$, when the temperature is slightly below the critical filling transition corresponding
to a value of the contact angle
 \bb
\theta-\frac{\pi}{4}\propto L^{-\frac{1}{\beta_w}}\,,\label{filling_shift}
 \ee
which quantifies the mesoscopic correction to the macroscopic phase diagram. For systems with dispersion or short-ranged intermolecular forces this
predicts $\theta-\frac{\pi}{4} \propto L^{-2}$ or $\theta-\frac{\pi}{4}\propto L^{-4}$, respectively. The finite-size scaling prediction
(\ref{filling_shift}) is reminiscent of the well-known scaling result for the shift of the interface localization-delocalization transition, below
the wetting temperature, in capillaries made from walls with competing wetting and drying properties \cite{parry90}.

\end{itemize}

\begin{figure}
\includegraphics[width=6cm]{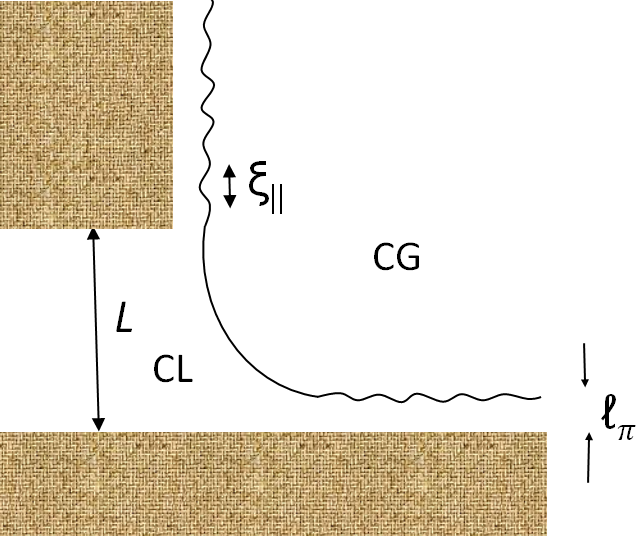}
\caption{Schematic illustration of the length-scales determining the mesoscopic rounding of the continuous meniscus depinning transition, due to
wetting layers adsorbed along the bottom and side walls. For example, for complete wetting, at a purely macroscopic level the meniscus rounds the
upper corner exactly when its radius $R=L$. The effective slit width however is altered by the wetting layer thickness and parallel correlation
length along the bottom and top walls, respectively. For systems with dispersion forces the meniscus depinning is rounded over a region $\Delta
p_{\rm md} \propto L^{-\frac{4}{3}}$. The rounding of the transition is sharper for partial wetting for which $\Delta p_{\rm md}\propto L^{-2}$.}
\label{meso}
\end{figure}

\item {\bf Rounding of the Meniscus Depinning Transitions}

At a macroscopic level, meniscus depinning is a continuous phase transition, which is of third-order phase  for complete wetting and second-order for
partial wetting. This, however, is rounded when we include mesoscopic length scales associated with wetting layers or if  the edges of the capillary
are no longer geometrically sharp. We consider rounding effects in turn beginning with case of complete wetting.

\noindent {\bf Complete wetting:} At a macroscopic level, it is clear that the location of meniscus depinning occurs when $R=L$, i.e, when we can
just fit a meniscus with a quarter circular shape into the ends of the capillary. However, this geometrical condition ignores the presence of the
complete wetting layers along the bottom and vertical walls, cf. Fig.~\ref{meso}. These are characterised by a thickness $\ell_{\pi}\approx \delta
p^{-\beta_s^{\rm co}}$ and also an parallel correlation length $\xi_\parallel \approx \delta p^{-\nu_{\parallel}^{\rm co}}$ arising from thermal
interfacial fluctuations \cite{dietrich}. Here $\beta_s^{\rm co}$ and $\nu_\parallel^{\rm co}$ are the critical exponents defined for the complete
wetting transition and take the values $\beta_s^{\rm co}=1/3$ and $\nu_\parallel^{\rm co}=2/3$ for systems with dispersion forces. The length scales
$\ell_\pi$ and $\xi_\parallel$ add uncertainty to the effective slit thickness arising from the wetting layers along the bottom and vertical walls
respectively with the contribution from the latter being dominant. Thus, we can expect that meniscus depinning occurs when $R\approx L\pm
\xi_\parallel$ or
 \begin{equation}
 \frac{\gamma}{\delta p}\approx L\left(1\pm \frac{\xi_\parallel}{L}\right)\,.
\end{equation}
It follows that the transition still occurs at $\delta p_{\rm md}=\gamma/L$ (equivalent to $\delta\tilde{p}_{\rm md}=1/2$) but is rounded over a
scale $\Delta p_{\rm md}\approx \xi_\parallel/L^2$. Allowing for the divergence of $\xi_\parallel$ at complete wetting this gives for the rounding of
the meniscus depinning when $\theta=0$,
 \begin{equation}
 \Delta p_{\rm md}\propto L^{\nu_\parallel^{\rm co}-2}\,.
 \end{equation}
This rounding of the meniscus depinning transition then determines that the change from type I to type II condensation is also not sharp but occurs
at $a_p=2/\pi$ rounded over a region
\begin{equation}
\Delta a_p\propto L^{\nu_\parallel^{\rm co}-1}\,.
\end{equation}
The additional factor of $L$ here arises because the $\delta \tilde{p}$--$a$ phase diagram is scale free.
For systems with dispersion forces this leads to the predictions $\Delta p_{\rm md} \propto L^{-\frac{4}{3}}$ and $\Delta a_p \propto
L^{-\frac{1}{3}}$ while for systems with strictly short-ranged forces the rounding of the meniscus depinning is sharper and occurs over the ranges
$\Delta p_{\rm md}\propto L^{-\frac{3}{2}}$ and $\Delta a_p \propto L^{-\frac{1}{2}}$.

We can now use the above finite-size scaling considerations to develop a crossover scaling theory for meniscus depinning transition for the case of
complete wetting. The purely macroscopic result (44) implies that, when we ignore the rounding due to complete wetting layers,  the grand potential
contains a singular contribution $\Omega_{\rm sing}=\gamma (R-L)^3/6L^2$. To allow for the rounding due to complete wetting layers, we anticipate
that this singular contribution is modified by a multiplicative scaling function $W(x)$. The argument of the scaling function $x$ must be
dimensionless and it is natural to identify this as $x=(\delta p-\delta p_{\rm md})/\Delta p_{\rm md}$ which is simply the relevant scaling field
divided by the predicted rounding.  Noting that $\delta p=\gamma/R$ and $\delta p_{\rm md}=\gamma/L$ it follows that the appropriate scaling ansatz
for the crossover and rounding of the meniscus depinning transition is
\begin{equation}
\Omega_{\rm sing}= \frac{(R-L)^3}{L^2}W \left(\frac{R-L}{L^{\nu_\parallel^{\rm co}}}\right)\,,
\end{equation}
where we have ignored all constants and metric factors highlighting only the dependence on the slit width.  We require that $W(x)\to0$ as
$x\to\infty$ and $W(x)\to 1$ as $x\to-\infty$ which represent the macroscopic unpinned and pinned states, respectively. The form of $W(x)$ describes
the smooth crossover between these two states when the mesoscopic wetting length-scale  $\xi_\parallel$ is allowed for. In particular, in order that
$\Omega_{\rm sing}$ is nonzero at the macroscopic meniscus depinning transition, $R=L$, we require that $W(x)\propto 1/x^3$, which leads to a
singular or mesoscopic contribution. Exactly at the predicted location of the macroscopic meniscus depinning transition, $R=L$, this crossover
scaling ansatz implies that the grand potential contains a singular contribution
\begin{equation}
\Omega_{\rm sing} \propto L^{3\nu_\parallel^{\rm co}-2}\,, \hspace{1cm} R=L\,.
\end{equation}

This may be regarded as a fluctuation-induced, Casimir-like, contribution to the free-energy and predicts that $\Omega_{\rm sing} \propto L^{-1/2}$
for systems with short-ranged forces. Intriguingly, for dispersion forces, for which $\nu_\parallel^{\rm co}=2/3$, the exponent vanishes which
probably corresponds to a marginal, logarithmic, contribution $\Omega_{\rm sing} \propto \ln L$. Indeed, this would be consistent with the known
logarithmic contribution to the finite-size excess free-energy of complete wetting drops in the presence of dispersion interactions \cite{stripe17}.
The derivative of $\Omega_{\rm sing}$ w.r.t. $\delta p$ determines the singular contribution to the adsorption, over and above the macroscopic
contribution. Since $\partial\Omega/\partial \delta p \propto R^2\partial\Omega/\partial R$ it follows that we can expect that the adsorption
contains a singular contribution
\begin{equation}
\Gamma_{\rm sing}= (R-L)^2\Lambda\left(\frac{R-L}{L^{\nu_\parallel^{\rm co}}}\right)\,,
\end{equation}
 where $\Lambda(x)$ is a suitable new scaling function trivially related
to $W(x)$. Again, we require that $\Lambda(x)\to0$ as $x\to\infty$ and $\Lambda(x)\to 1$ as $x\to-\infty$ and $\Lambda(x)\propto 1/x^2$ as $x\to0$.
It follows that exactly at the (macroscopic) depinning phase boundary, $R=L$, the excess adsorption contains a fluctuation-induced contribution
\begin{equation}
\Gamma_{\rm sing} \propto L^{2\nu_\parallel^{\rm co}}\,, \hspace{1cm} R=L\,, \label{gamma_sing_comp}
\end{equation}
in addition to the leading-order macroscopic term $\Gamma=2\Delta \rho \left(1-\frac{\pi}{4}\right) L^2$ determined earlier. Next, we recall the
exact exponent relation for complete wetting, $2\nu_\parallel^{\rm co}=1+\beta_s^{\rm co}$ (see Ref.~\cite{schick}) so that we can also identify this
mesoscopic contribution as $\Gamma_{\rm sing} \propto L^{1+\beta_s^{\rm co}}$. The physical meaning of this contribution is now apparent since the
factor $L^{\beta_s^{\rm co}}$ is simply the thickness of the complete wetting layer $\ell_\pi \propto \delta p^{-\beta_s^{\rm co}}$ evaluated at the
pressure of the meniscus depinning transition $\delta p_{\rm md}=\gamma/L$. Thus, the mesoscopic term can be written
\begin{equation}
\Gamma_{\rm sing}\propto \ell_\pi L\,, \hspace{1cm} R=L\,,
\end{equation}
which, of course, is simply the additional contribution to the adsorption from the meniscus when we shift its position by the thickness of a wetting
layer coating the side walls. Explicitly, for systems with dispersion forces this yields $\Gamma_{\rm sing} \propto L^{4/3}$.

\noindent {\bf Critical wetting:} There is a very simple extension of this crossover scaling theory to the case of critical wetting when we suppose
that the meniscus depinning occurs exactly at the temperature $T_w$ of a continuous wetting transition. This also corresponds to a case where
$\theta=0$ and that the parallel correlation length $\xi_\parallel$ would diverge as $\delta p$ tends to zero. The only difference with the analysis
of complete wetting is that in place of the expression $\xi_\parallel \approx \delta p^{-\nu_\parallel^{\rm co}}$ we must use $\xi_\parallel \approx
\delta p^{-\nu_\parallel/\Delta_s}$ as is appropriate for the divergence of the parallel correlation length along the critical isotherm of the
continuous (critical) wetting transition. Here $\nu_\parallel$ and $\Delta_s$ are the correlation length and gap exponents for the critical wetting
transition. For meniscus depinning occurring exactly at  critical wetting the adsorption therefore contains a mesoscopic contribution $\Gamma_{\rm
sing} \propto L^{2\nu_\parallel/\Delta_s}$. Again, this can be interpreted as simply $\Gamma_{\rm sing}\propto \ell_\pi L$ where
$\ell_\pi\propto\delta p_{\rm md}^{-\beta_s/\Delta_s}$ is the thickness of the wetting layer.  More explicitly, for systems with dispersion forces
this yields $\Gamma_{\rm sing} \propto L^{5/4}$ in contrast to the $L^{4/3}$ power law for complete wetting. We make these remarks, not because this
scenario is likely to be observed, but because it will consistently tie in with the crossover scaling theory for the case of partial wetting which we
develop next.

\noindent {\bf Partial wetting:} The same considerations apply to the rounding of the meniscus depinning and type I/II crossover in the partial
wetting but complete filling regime $0<\theta<\pi/4$. However, there is a difference in the quantitative nature of the rounding because the parallel
correlation length $\xi_\parallel$ remains finite as saturation is approached. Nevertheless, this is still the relevant microscopic length scale
determining the rounding of the transitions leading to the universal finite-size scaling predictions
\begin{equation}
\Delta p_{\rm md}\propto L^{-2}\,;\;\;\; (0<\theta<\pi/4) \label{delta_p_part}
\end{equation}
and
\begin{equation}
\Delta a_p\propto L^{-1}\,;\;\;\; (0<\theta<\pi/4)\,, \label{delta_a_part}
\end{equation}
for the meniscus depinning and type I/II crossover, respectively. Therefore both the meniscus depinning and type I/II crossover is significantly
sharper for partial wetting due to the absence of the complete wetting layers.

Following a similar line of argument to our discussion of complete (and critical) wetting we can now develop a crossover scaling theory for the
rounding of the meniscus depinning for the case of partial wetting where, at a macroscopic level, the transition is second-order. We do this directly
for the singular contribution to the adsorption and also for small contact angles which will allow us to connect with the limit of complete wetting.
From Eq.~(48) it follows that at a macroscopic level the second-order meniscus depinning transition is associated with a singular contribution
$\Gamma_{\rm sing} \propto \theta L (R- cL)$ where $c=1/(1-\theta)$ follows from (45) and characterises the shift in the location of the transition
when the contact angle is finite.  To allow for the rounding of the phase transition due to the partial wetting layers, we multiple this by a
suitable scaling function $\tilde \Lambda(x)$ where again we may identify the dimensionless scaling variable $x=(\delta p-\delta p_{\rm md})/\Delta
p_{\rm md}$ with $\Delta p_{\rm md} \propto \xi_\parallel/L^2$. This immediately determines that the crossover from a pinned to unpinned
configuration is associated with a mesoscopic contribution
\begin{equation}
\Gamma_{\rm sing} = \theta L (R- cL) \tilde \Lambda\left(\frac{R - cL}{\xi_\parallel}\right)\,.
\end{equation}
For partial wetting, we require that the scaling function has the (macroscopic) limits $\tilde \Lambda(x)\to0$ as $x\to\infty$ and  $\tilde
\Lambda(x)\to1$ as $x\to-\infty$ together with the (continuity) condition $\tilde \Lambda(x)\propto 1/x$ as $x\to0$. For completion we note that the
corresponding scaling ansatz for the grand potential is
 \bb
 \Omega_{\rm sing}= \frac{\theta}{L}(R-cL)^2\tilde W\left(\frac{R - cL}{\xi_\parallel}\right)\,,
 \ee
where $\tilde W(x)$ is a suitable scaling function with similar macroscopic and continuity properties. As we shall see, this can be interpreted as a
contribution arising from the line tension. Exactly at the pressure of the macroscopic meniscus depinning transition ($\delta p=\delta p_{\rm md}$,
equivalent to $R=cL$), this determines that there is a mesoscopic contribution to the adsorption
\begin{equation}
\Gamma_{\rm sing} \propto \theta \xi_\parallel L\,, \hspace{1cm} R=cL\,, \label{gamma_sing_part}
\end{equation}
in addition to the macroscopic term $\Gamma=2\Delta \rho L^2 (1-\frac{\pi}{4})/(1-\theta)^2$, which follows directly from (47) when $\theta$ is
small. This mesoscopic contribution is analytic in $L$, as may be anticipated, since there are no diverging mesoscopic length scales. Nevertheless,
this simple result consistently explains how a non-trivial power-law dependence on $L$ emerges when we consider the limit $\theta\to 0$ corresponding
to the approach to a critical wetting transition. To see this, we recall that the critical wetting exponent relation
$2-\alpha_s=2\nu_\parallel-2\beta_s$ (involving the standard critical exponents for the surface specific heat, adsorption, and parallel correlation
length -- see Ref.~\cite{schick} for details) is equivalent to identifying the contact angle $\theta \propto \ell_\pi/\xi_\parallel$. In other words,
as with the result (60) for complete wetting, we can interpret this mesoscopic contribution to the adsorption at meniscus depinning, as
\begin{equation}
\Gamma_{\rm sing}\propto \ell_\pi L, \hspace{1cm}  R=cL\,,
\end{equation}
arising directly from the shifted position of the menisci due to the wetting layer. This contribution is analytic for partial wetting since
$\ell_\pi$ remains microscopic. However, on approach a critical wetting transition we must insert $\ell_\pi \propto L^{\beta_s/\Delta_s}$ which
recovers consistently the result $\Gamma_{\rm sing} \propto L^{2\nu_\parallel/\Delta_s}$ derived earlier for the case $\theta=0$.

A similar physical interpretation applies to the mesoscopic contribution to the grand potential for partial wetting. Exactly at the macroscopic
meniscus depinning phase boundary, $R=cL$, the scaling ansatz implies that $\Omega_{\rm sing}\propto \theta \xi_\parallel^2/L$ and hence $\Omega_{\rm
sing}\propto \ell_\pi\xi_\parallel/L$  We now allow for the critical singularities of the wetting film thickness, $\ell_\pi \propto \tilde
t^{-\beta_s}A(\delta p \tilde t^{-\Delta_s})$, and parallel correlation length, $\xi_\parallel \propto \tilde t^{-\nu_\parallel} B(\delta p \tilde
t^{-\Delta_s})$, in the vicinity of a critical wetting transition with $\tilde t=(T_w-T)/T$. Note that this necessarily includes the behaviour of
these lengthscales off bulk coexistence which are described by scaling functions $A(\cdot)$ and $B(\cdot)$ with $\delta p$ evaluated at the location
of the meniscus depinning transition, $\delta p_{\rm md} \propto 1/L$. Substituting for $\ell_\pi$ and $\xi_\parallel$, and then simply multiplying
numerator and denominator by $\tilde t^{\Delta_s}$, we observe that the singular contribution to the grand potential reduces to $\Omega_{\rm sing}
\propto \tilde t^{-\beta_s-\nu_\parallel+\Delta_s} C(L \tilde t^{\Delta_s})$ with $C(y)$ a suitable function of the scaling variable $y=L \tilde
t^{\Delta_s}$. Finally, we use the standard critical exponent relation $2-\alpha_s-\Delta_s=-\beta_s$ (again, see Ref.~\cite{schick}), to get our
desired result
\begin{equation}
\Omega_{\rm sing} \propto \tilde t^{2-\alpha_l} C(L \tilde t^{\Delta_s})\,.
 \end{equation}
Here $\alpha_l=\alpha_s+\nu_\parallel$ is nothing other than the critical exponent characterizing the singularity in the line tension, $\tau_l\propto
\tilde{t}^{2-\alpha_l}$, on approaching the wetting temperature $T_w$ \cite{indekeu}. In other words, the present crossover scaling theory for the
rounding of the macroscopic meniscus depinning transition is equivalent to allowing for the line tension associated with the contact of the meniscus
with the corner and walls. We note that the scaling function $C(y)$ must satisfy $C(\infty)=1$ and $C(y) \propto y^{(\alpha_l-2)/\Delta_s}$ as $y \to
0$, to ensure that this mesoscopic contribution to the grand potential exists away from and at the critical wetting transition itself. Together with
the direct, physically intuitive, interpretation of the mesoscopic contribution to the adsorption, we regard this as convincing support for the
crossover scaling theory for the rounding of the depinning transition.

\noindent {\bf Wall structure and roughness:} The meniscus depinning transition and type I/II crossover will also be rounded if the edge of the
capillary is no longer geometrically sharp -- for example, if the upper corners of the slit are not modelled as perfect right angles but instead as
quarter circles of radius $r_e$, say. This will always be the case at a microscopic level and presumably, for real solids, the smallest value of
$r_e$ corresponds to a few molecular diameters $\sigma$. The length scale $r_e$ also serves to round the meniscus depinning transition, replacing
$\xi_\parallel$ in the argument given above, giving rise to similar predictions (\ref{delta_p_part}) and (\ref{delta_a_part}), but now induced by an
underlying geometrical roughness. Thus, we anticipate that the meniscus depinning transition is always rounded, either by the thermal fluctuations of
wetting layers or wall roughness or molecular structure. For complete wetting, when $L$ is large, the rounding is dominated by the effects arising
from interfacial wandering, characterised by $\xi_\parallel$, while for partial wetting, the effects of wetting layers and wall molecular structure
are comparable. We anticipate that the minimum value of the uncertainty in the location of type I/II crossover is $\Delta a_p \approx \sigma/L$,
which will be significant for nanoscopic slits.

\end{enumerate}


\section{Density functional theory}

In this section we compare our predictions with a microscopic DFT model \cite{evans79}, which will allow us to study these phenomena at the molecular
scale. We concentrate on two aspects: firstly, that for complete wetting, the capillary condensation occurs over the whole range of accessible aspect
ratios and is of type I for small $a$ and type II for large $a$. Secondly, in the partial filling regime, $\theta>\pi/4$ condensation is only of type
I and is suppressed for sufficiently large aspect ratios $a>a_0$, which we determine and compare with the theoretical prediction $a_0=\cot\theta$.

To this end, we employ the same DFT model that we have used recently for the $HH$ geometry, which combines Rosenfeld's fundamental measure theory
\cite{ros} describing accurately any packing effects, with a mean-field treatment of the attractive part of the inter-atomic interaction modelled by
a truncated Lennard-Jones potential, see e.g. \cite{mal13, laska21} for explicit details. The mean-field DFT model misses some fluctuation effects
associated with interfacial wandering and the rounding of the capillary condensation transition; however, these play no role in determining the
location of the type I and type II capillary condensation and meniscus depinning transitions which are of our central concern here.

Within classical DFT, the equilibrium one-particle density $\rhor$ of an inhomogeneous fluid is determined by minimization of the grand potential
functional
 \begin{equation}
 \Omega[\rho]={\cal F}[\rho]+\int\dd\rr\rhor[V(\rr)-\mu]\,.
\end{equation}
Here, ${\cal F}[\rho]$ is the intrinsic free-energy functional which contains all the information about the model fluid, $V(\rr)$ is the external
potential which, in our case, represents the effect of the confining walls and $\mu$ is the chemical potential. The intrinsic free-energy functional
can be separated into  an ideal-gas term, ${\cal F}_{\rm id}$, and an excess part, ${\cal F}_{\rm ex}$, arising from the fluid-fluid interaction:
 \bb
 {\cal F}[\rho]={\cal F}_{\rm id}[\rho]+{\cal F}_{\rm ex}[\rho]\,.
 \ee
 The ideal-gas term due to purely entropic effects is known exactly:
     \bb
  \beta F_{\rm id}[\rho]=\int\dr\rho(\rr)\left[\ln(\rhor\Lambda^3)-1\right]\,,
  \ee
  where $\Lambda$ is the thermal de Broglie wavelength and $\beta=1/k_BT$ is the inverse temperature.

The fluid is modelled using a truncated (and non-shifted) Lennard-Jones potential, in which case the excess contribution can be treated in a
perturbative manner and is further split into a contribution ${\cal F}_{\rm hs}$ due to short-range repulsive forces approximated by a hard-sphere
potential, and a contribution ${\cal F}_{\rm att}$ arising from the attractive interactions:
 \bb
{\cal F}_{\rm ex}[\rho]={\cal F}_{\rm hs}[\rho]+{\cal F}_{\rm att}[\rho]\,.
 \ee
The repulsion part of the free-energy is described using Rosenfeld's fundamental measure theory  \cite{ros}
 \bb
{\cal F}_{\rm hs}[\rho]=k_BT\int\dd\rr\,\Phi(\{n_\alpha\})\,,
 \ee
where  the free energy density $\Phi$ depends on the set of weighted densities $\{n_\alpha\}$. Within the original Rosenfeld approach these consist
of four scalar and two vector functions, which are given by convolutions of the density profile and the corresponding weight function:
 \bb
 n_\alpha(\rr)=\int\dr'\rho(\rr')w_\alpha(\rr-\rr')\;\;\alpha=\{0,1,2,3,v1,v2\}\,,
 \ee
where $w_3(\rr)=\Theta(R-|\rr|)$, $w_2(\rr)=\delta(R-|\rr|)$, $w_1(\rr)=w_2(\rr)/4\pi R$, $w_0(\rr)=w_2(\rr)/4\pi R^2$,
$w_{v2}(\rr)=\rr\delta(R-|\rr|)/R$, and $w_{v1}(\rr)=w_{v2}(\rr)/4\pi R$. Here, $\Theta$ is the Heaviside function and the hard-sphere radius is set
to $R=\sigma/2$ where $\sigma$ is the fluid potential parameter defined below.

The attractive free-energy contribution is treated at a mean-field level:
 \bb
 F_{\rm att}[\rho]=\frac{1}{2}\int d{\bf{r}}_1\rho(\rr_1)\int d{\bf{r}}_2\rho(\rr_2)u_{\rm att}(|\rr_1-\rr_2|)\,,
 \ee
 where $u_{\rm att}(r)$ is the attractive part of the Lennard-Jones-like potential
 \bb
 u_{\rm a}(r)=\left\{\begin{array}{cc}
 0\,;&r<\sigma\,,\\
-4\varepsilon\left(\frac{\sigma}{r}\right)^6\,;& \sigma<r<r_c\,,\\
0\,;&r>r_c\,.
\end{array}\right.\label{ua}
 \ee
which is truncated at $r_c=2.5\,\sigma$. For this model, the critical temperature corresponds to $k_BT_c=1.414\,\varepsilon$.

We begin by considering the phase diagram for the case of complete wetting. Actually, we flip the scenario and consider walls which have a purely
long-ranged repulsive component which ensures that the horizontal and vertical surfaces are all completely dry with contact angle $\theta=\pi$. The
phase diagram shown in Fig.~5a remains unchanged except that now we consider the reservoir to be a dense liquid and that the fluid in the slit
undergoes capillary evaporation as the pressure is \emph{reduced} to $p_{\rm sat}$ (that is the roles played by the CG and CL phases are simply
reversed in Fig.~5a). By focussing on drying we also avoid the aforementioned issues related to molecular layering and volume exclusion. Finally, we
add that we use a long-ranged repulsion instead of a pure hard-wall to better model and numerically handle the corner edges of the $H\infty$
geometry.

\begin{figure}
\includegraphics[width=8cm]{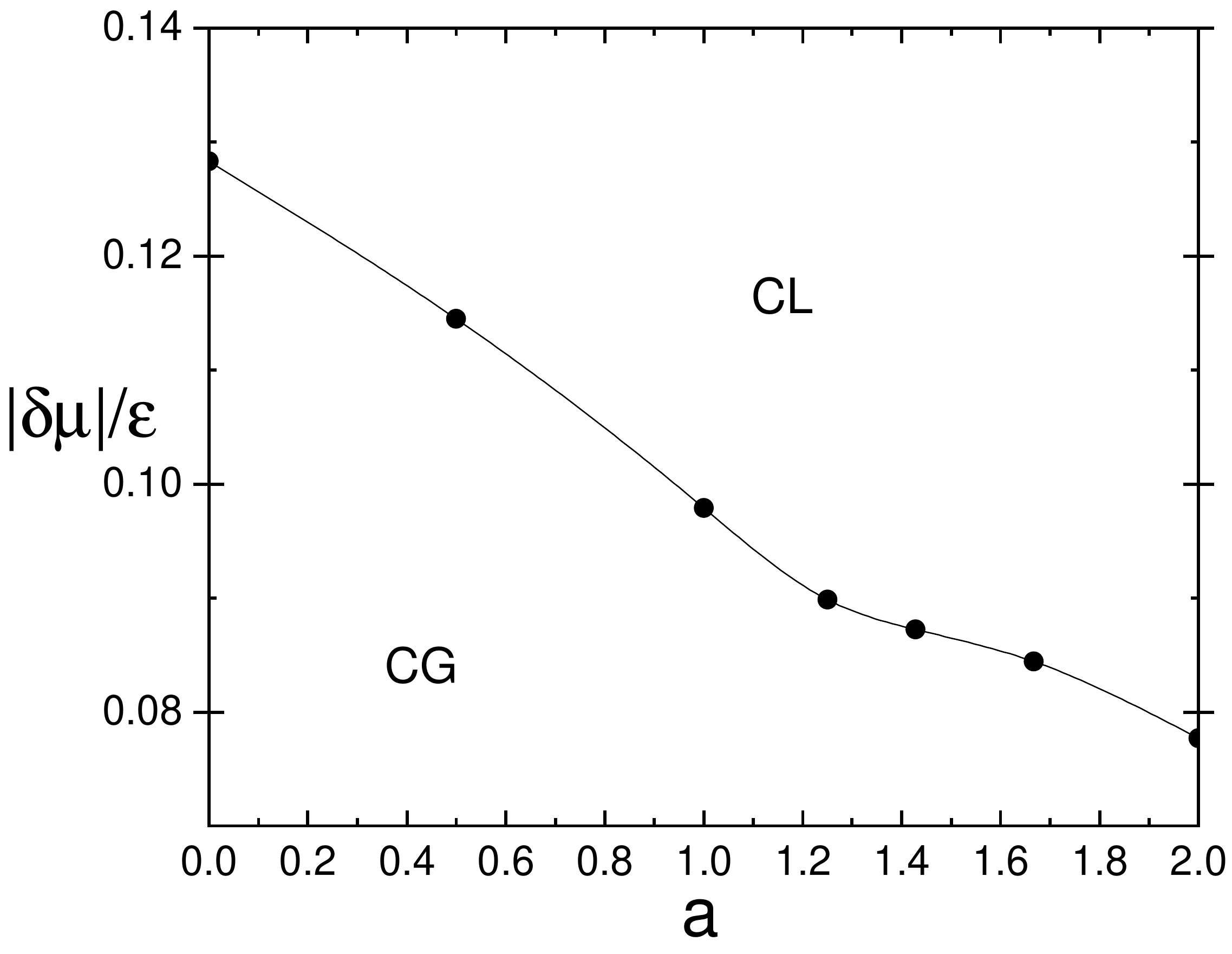}
\caption{Microscopic DFT results for the phase diagram in an $H\infty$ open slit with purely repulsive walls, corresponding to complete drying
$\theta=\pi$, showing the locus of capillary evaporation as a function of increasing aspect ratio. Here the chemical potential shift from saturation
is measured in units of the strength of the fluid-fluid potential. Capillary evaporation occurs for all values of the aspect ratio.} \label{fig9}
\end{figure}

The repulsive walls are assumed to be formed of atoms distributed uniformly with a number density $\rho_w$ which interact with the fluid atoms via
the repulsive part of the Lennard-Jones potential, $\phi^r(r)=4\varepsilon_w\left(\frac{\sigma_w}{r}\right)^{12}$. The net potential induced by the
walls can be split into a term $V_{\rm bottom}^r$ due to the bottom (planar) wall and the contribution $V_{\rm top}^r$ of the top wall of width $H$.
Both are determined by integrating $\phi^r(r)$ over the whole domains of the respective walls:
 \bb
 V^r(L,H; x,z)=V_{\rm bottom}^r(z)+V_{\rm top}^r(H;L-x,z)
 \ee
 where $V_{\rm bottom}^r(z)=\frac{4}{45}\pi\varepsilon_w\rho_w\sigma^3\left(\sigma/z\right)^9$ and
 \begin{eqnarray}
V_{\rm top}^r(H;x,z)&=&\pi\varepsilon_w\sigma^{12}\rho_w\left[\psi_{12}(x,\infty)-\psi_{12}(x,z)\right.\\
&&\left.-\psi_{12}(x-H,\infty)+\psi_{12}(x-H,z)\right]\nonumber
 \end{eqnarray}
 with
\begin{widetext}
 \bb
 \psi_{12}(x,z)=-\frac{1}{2880}\frac {128\,{x}^{16}+448\,{x}^{14}{z}^{2}+560\,{x}^{12}{z}^{4}+280\,{
x}^{10}{z}^{6}+35\,{x}^{8}{z}^{8}+280\,{x}^{6}{z}^{10}+560\,{x}^{4}{z} ^{12}+448\,{z}^{14}{x}^{2}+128\,{z}^{16}}{{z}^{9}{x}^{9} \left( {x}^{2
}+{z}^{2} \right) ^{7/2}}\,.
 \ee
 \end{widetext}

 \begin{figure}

\vspace*{0.5cm}

\includegraphics[width=7cm]{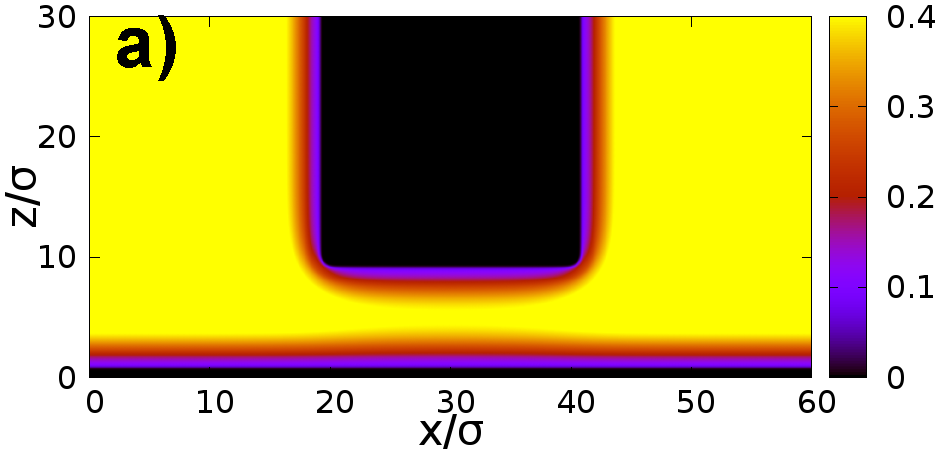}

\vspace*{0.5cm}

\includegraphics[width=7cm]{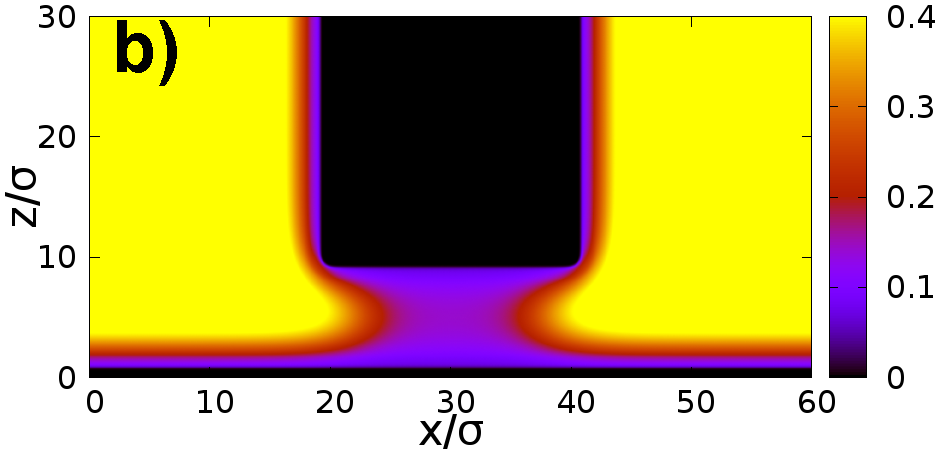}

\vspace*{0.5cm}

\includegraphics[width=7cm]{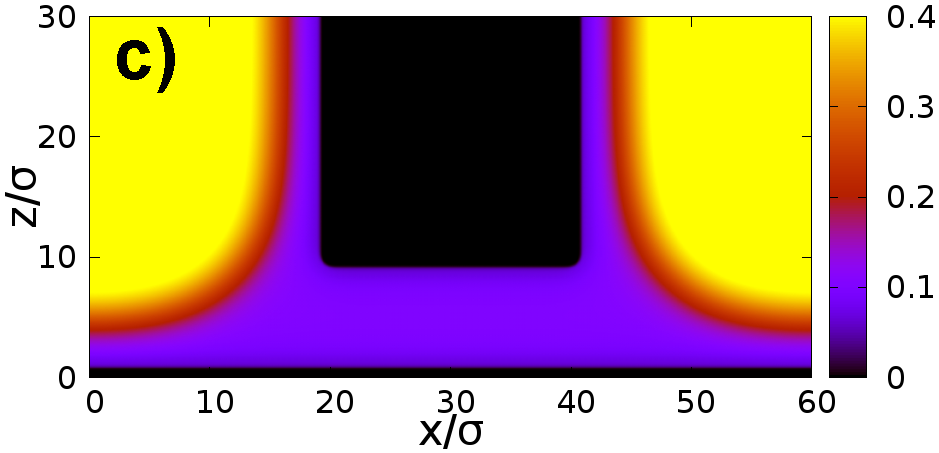}
\caption{Numerically determined DFT results for the density profiles for an $H\infty$ slit with repulsive walls corresponding to $\theta=\pi$ for a
long slit with aspect ratio $a=1/2$. The plots a) and b) show the coexisting CL and CG phases, respectively, where the menisci for the latter
evaporated phase are clearly pinned and located within the slit. The plot c) shows the density profile closer to saturation where it is clear that
the menisci are unpinned and are located outside the slit.} \label{dens_profs_H20}
\end{figure}

\begin{figure}

\vspace*{0.5cm}

\includegraphics[width=7cm]{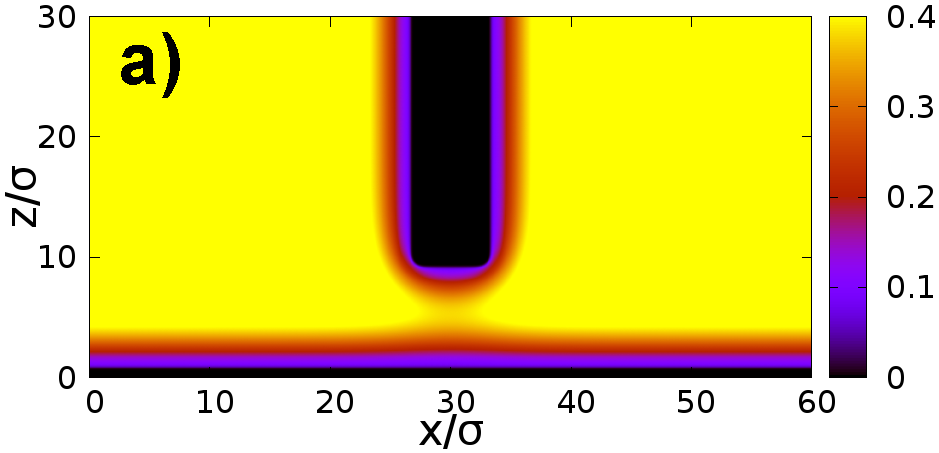}

\vspace*{0.5cm}

\includegraphics[width=7cm]{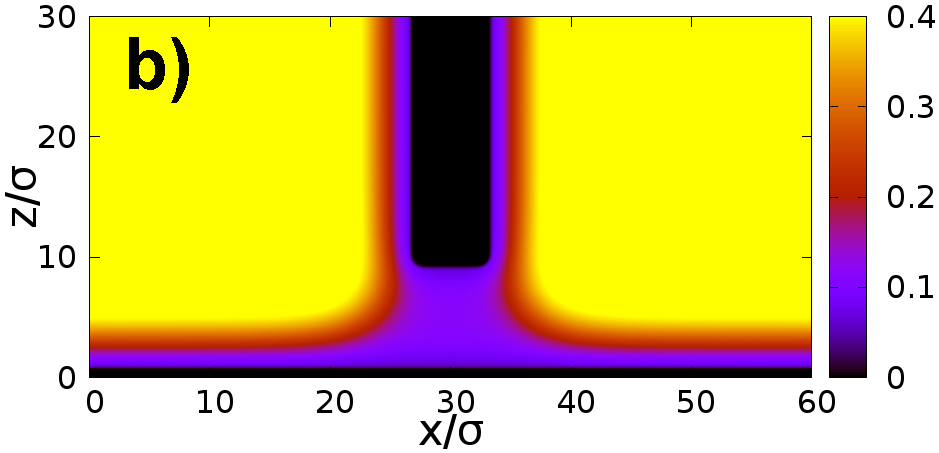}

\caption{Numerically determined DFT results for the density profiles for an $H\infty$ slit with repulsive walls corresponding to $\theta=\pi$ for a
short slit with aspect ratio $a=1$. the plots show the coexisting CL (a) and CG (b) phases where the menisci for the latter evaporated phase are
unpinned and located outside the slit. }\label{dens_profs_H5}
\end{figure}

In Fig.~\ref{fig9} we show the phase diagram obtained from the DFT at $T=0.92\,T_c$, where $T_c$ is the bulk critical temperature of the model fluid,
and $\rho_w\varepsilon_w=0.8\,\varepsilon\sigma^{-3}$. We use the fluid-fluid Lennard-Jones potential parameters $\sigma$ and $\varepsilon$ as the
appropriate units for length and energy, respectively. Here, the vertical axis is expressed in terms of the chemical potential difference from its
value at saturation $\delta\mu=\mu-\mu_{\rm sat}$, which we recall is related to the pressure difference approximately as $\delta p\approx \delta\mu
\Delta\rho$. We have obtained the phase diagram showing the line of capillary condensation over a wide range of the aspect ratio for a microscopic
slit separation $L=10\,\sigma$, by computing the loci where the CG and CL phases have the same numerically determined equilibrium grand potential.
Our results are consistent with the predicted shape of the macroscopic phase diagram and also demonstrate clearly that the condensation involve
pinned and unpinned menisci for small and large aspect ratios, respectively. Figs.~\ref{dens_profs_H20} a,b  show the coexisting CL and CG phases for
aspect ratio $a=1/2$ in which the pinning of the menisci at the ends is clearly visible. On further reducing the chemical potential towards
saturation these menisci depin and round the corners as shown in Fig.~\ref{dens_profs_H20}c. However, when we reduce the value of $H$, the character
of the capillary condensation changes. This is illustrated in Fig.~\ref{dens_profs_H5} which shows the coexisting CL and CG phases for an aspect
ratio $a=2$ where the menisci at capillary evaporation lie outside the capillary slit. The two values of $a$ chosen here lie either side of the
predicted value $a_p=2/\pi$. However, the microscopic size of this capillary means that the crossover from type I to type II condensation is smooth.

For the second part of our DFT study we test the predicted form of the phase diagram Fig.~5d in the partial filling regime. To this end we must add
an attractive part to the substrate-fluid potential in order to decrease the contact angle. We also return to the original scenario where the
reservoir is a bulk gas and consider the capillary condensation that occurs when the chemical potential is increased towards saturation. We assume
the walls are made of atoms interacting with the fluid via the Lennard-Jones $12$-$6$ potential, in which case the potential of the bottom wall
becomes the familiar Lennard-Jones $9$-$3$ potential:
 \bb
 V_{\rm bottom}(z)=4\pi\varepsilon_w\rho_w\sigma^3\left[\frac{1}{45}\left(\frac{\sigma}{z}\right)^9-\frac{1}{6}\left(\frac{\sigma}{z}\right)^3\right]\,.
 \ee
The potential of the top wall will now be of the form:
 \bb
V_{\rm top}(H;x,z)=V_{\rm top}^r(H;x,z)+V_{\rm top}^a(H;x,z)\,,
 \ee
 where the attractive portion of the potential is
  \begin{eqnarray}
 V_{\rm top}^a(H;x,z)&=&\alpha_w\left[\frac{1}{(H-x)^3}+\psi_6(x-H,z)\right.\nonumber\\
 &&\left.-\frac{1}{x^3}-\psi_6(x,z)\right]\,,
  \end{eqnarray}
  with
  \bb
  \alpha_w=-\frac{1}{3}\pi\varepsilon_w\sigma^6\rho_w
 \ee
 and
 \bb
 \psi_6(x,z)=-{\frac {2\,{x}^{4}+{x}^{2}{z}^{2}+2\,{z}^{4}}{2{z}^{3}{x}^{3}
\sqrt {{x}^{2}+{z}^{2}}}}\,.
  \ee

  \begin{figure}
\includegraphics[width=8cm]{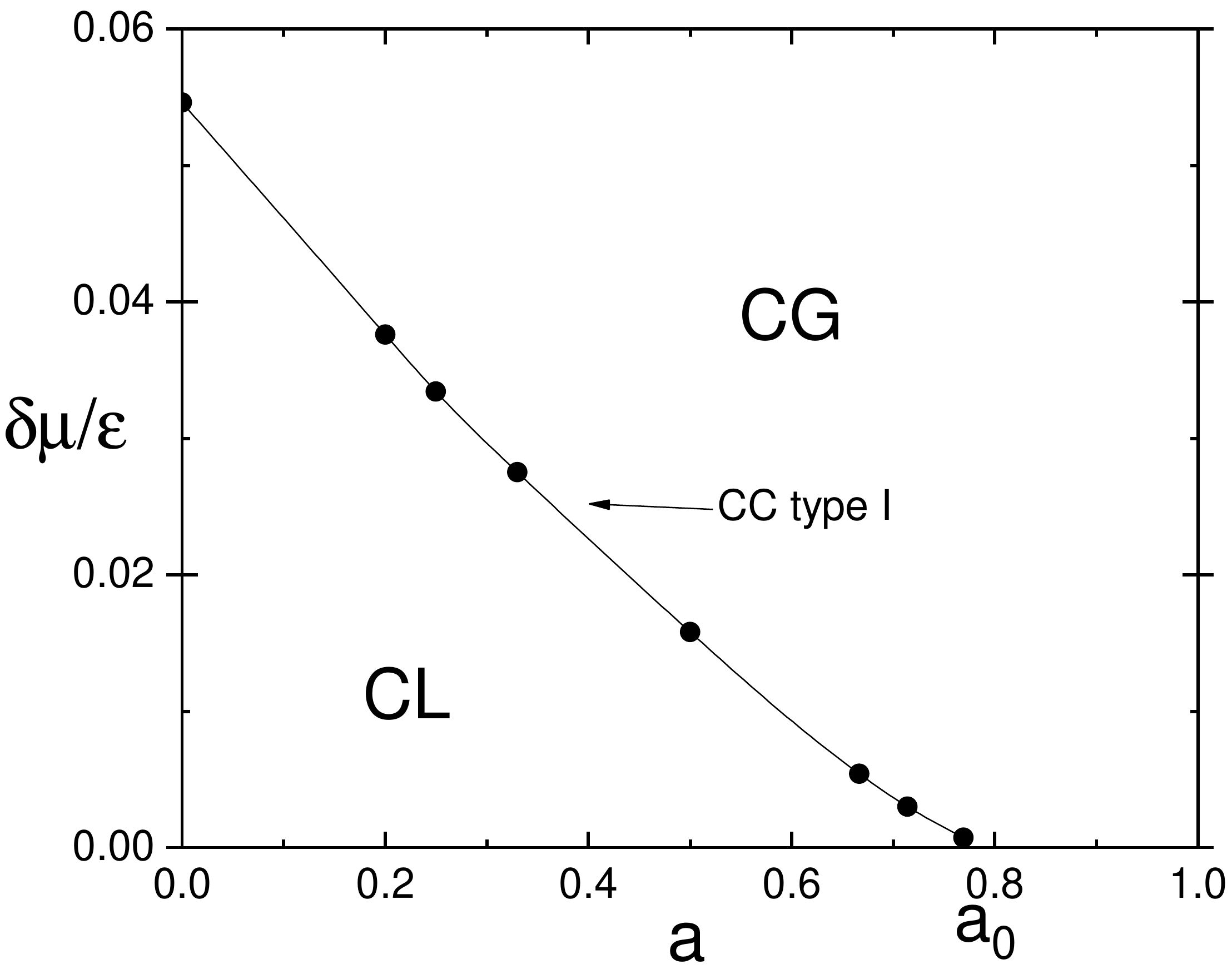}
\caption{Numerically determined DFT results for the phase diagram for an $H\infty$ slit for walls with a full Lennard-Jones potential, which would
give rise to partial corner filling with contact angle $\theta\approx 53\degree$. Capillary condensation is only of type I and ends, at bulk
saturation, when the aspect ratio $a_0\approx 0.78$, which is very to the theoretical prediction $a_0\approx 0.75$ given by Eq.~(\ref{a0}).}
\label{fig_dft_t12}
\end{figure}

\begin{figure}

\vspace*{0.5cm}

\includegraphics[width=7cm]{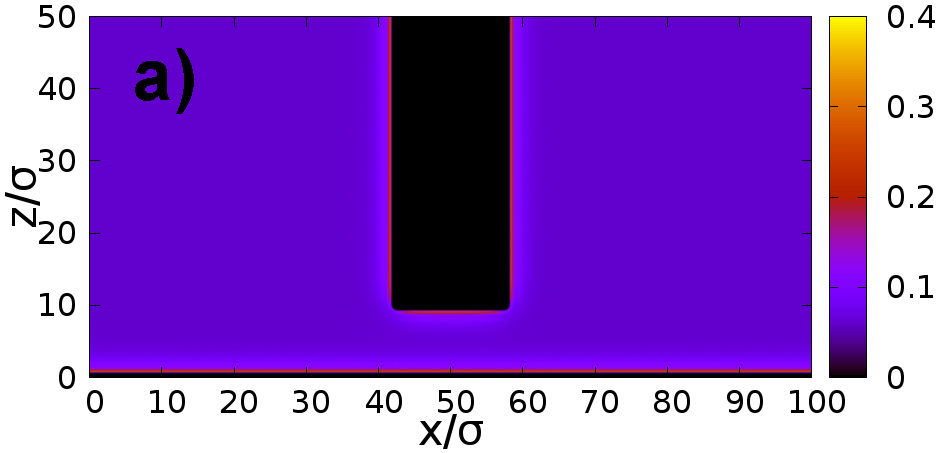}

\vspace*{0.5cm}

\includegraphics[width=7cm]{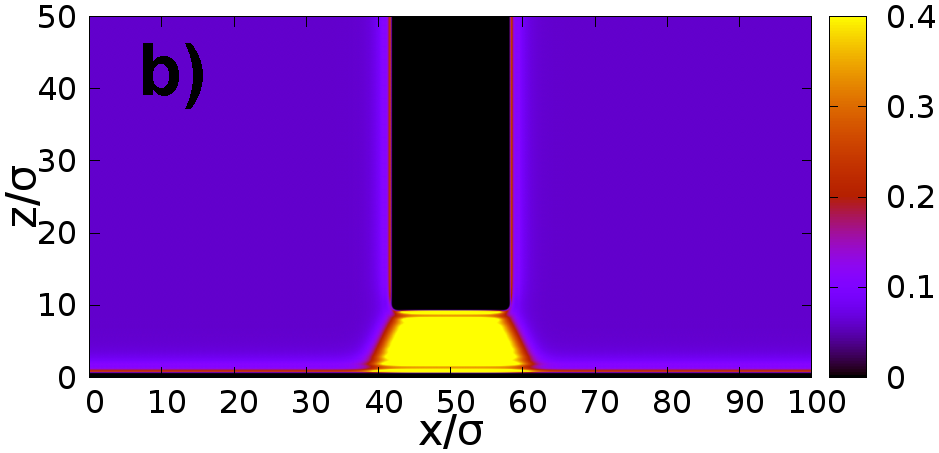}
\caption{Numerically determined DFT results for the coexisting density profiles for type I capillary condensation in a short slit with aspect ratio
$a=2/3$ for walls with  partial wetting $\theta=53\degree$. The menisci at capillary evaporation, which separate the capillary liquid from the
outside gas reservoir, are pinned, and lie within the slit. These menisci are near planar since the capillary condensation occurs close to bulk
saturation.} \label{dens_profs_H15}
\end{figure}

The wetting temperature of a planar wall corresponding to this full Lennard-Jones potential is known to be $T_w\approx0.91\,T_c$ \cite{bridge}. We
set the temperature  $T=0.85\,T_c$, for which the Young contact angle is $\theta\approx53\degree$ which therefore lies in the partial filling regime.
The numerically determined phase diagram showing the line of capillary condensation which terminates at bulk coexistence at a specific value of $a_0$
is displayed in Fig.~\ref{fig_dft_t12}. We observed only type I capillary condensation involving pinned menisci  over the whole range of aspect
ratios up to the maximal value $a_0\approx0.78$ at which the condensation occurs at bulk coexistence.  Representative density profiles of the
coexisting states (for $H=15\,\sigma$) for which $a=2/3$ are shown in Fig.~\ref{dens_profs_H15}. Our numerically determined value for $a_0$ is
extremely close to the macroscopic theoretical prediction $a_0\approx0.75$ given   Eq.~(\ref{a0}).

\section{Summary}

In this paper we have considered the phase equilibria of a fluid in an open slit formed when a wall of finite length $H$ is brought near a substrate
of infinite extent. In the first part of our paper we focused on understanding the basic macroscopic aspects of the possible phase behaviour which is
now much richer than that for the related $HH$ geometry in which both walls are the same length. This richness emerges because each aspect of the
$H\infty$ geometry brings with it the possibility of a phase transition. Thus, the proximity of the two parallel walls may induce capillary
condensation and the openness of the slit ends means that the condensed phase must involve menisci. The presence of edges mean that the upper parts
of the menisci may be pinned at these upper corners while the macroscopic extent of the lower wall mean that the menisci may overspill into the
right-angle corners and be unpinned. The fact the resulting phase diagram shows two types of capillary condensation, involving either pinned (type I)
or unpinned (type II) menisci, arises directly from the resulting marriage with the meniscus depinning and corner filling transitions. Finally,
condensation can always be suppressed by reducing the length of the slit when the free-energy cost of creating menisci becomes too great. At the
point where the type I condensation is suppressed the menisci are flat and the edge contact angle $\theta_e^{\rm cc}=\pi-\theta$. Therefore, at this
point, the capillary condensation mimics the phase separation occurring in infinite slit with materially different walls with opposing wetting
properties.

Our macroscopic results are summarised in the phase diagrams shown in Fig.~5, which shows the ($\delta\tilde{p}$-$a$) section for different $\theta$
and Fig.~6 which shows the ($\delta\tilde{p}$-$\theta$) section for different aspect ratio. The portrayal of the possible phase equilibria in terms
of ($\delta\tilde{p}$-$\theta$) is perhaps most physically relevant as, in practice, it is easier to continuously vary the contact angle rather than
the aspect ratio. 
The macroscopic phase diagram falls into three possible regimes
delineated by universal values of the aspect ratio. {\it Long} capillaries, with $a<2/\pi$, for which the condensation is always of type I and there
is a separate third-order continuous meniscus depinning transition. {\it Intermediate} capillaries, with $2/\pi<a<1$, where the condensation is
either of type I (for $\theta>\theta_p$) or type II (for $\theta<\theta_p$) where the separatrix between these two regimes intersects the line of the
meniscus depinning transition. And finally, {\it short} capillaries, with $a>1$, for which only type II condensation exists and there is no meniscus
depinning. In each of these three regimes condensation is suppressed for sufficiently large values of the contact angle $\theta>\theta_0$, the value
of which depends on $a$.

We have shown that, at macroscopic level, meniscus depinning is a continuous phase transition which is third-order for complete wetting and
second-order for partial wetting; to the best of our knowledge this is a new example of an interfacial phase transition, in which the first or second
derivative of the adsorption is discontinuous.   Meniscus depinning does not involve the divergence of the adsorption itself as in wetting and
wedge-filling transitions, nor does it involve the coexistence of different phases as in first-order wetting and prewetting; at the meniscus
depinning transition the pinned and depinned states are identical. We have discussed the rounding of meniscus depinning transitions which occur on
mesoscopic level due to the presence of wetting layers using the crossover scaling theory which allows for the direct influence of intermolecular
forces and thermal interfacial fluctuations. Similar rounding will occur if the edge of the capillary slit is not geometrically sharp which of course
will always be the case on the molecular level. However, even allowing for these, there is essentially no rounding of the suppression of capillary
condensation as the aspect ratio is increased (for a given value of $\theta$). Indeed, in mean-field studies this suppression of capillary
condensation remains a sharp effect. This we have illustrated using a microscopic DFT model which shows that the macroscopic prediction for the value
of $a_0$ is extraordinarily accurate down to the molecular scale.

We can extend the present study in several ways by, for example, supposing that the bottom wall is of finite length, but still longer than the top
wall, say. Depending on the extension of the bottom wall, the meniscus may also be either pinned or unpinned at this lower corner, similar but not
quite identical to the pinning discussed here for the top corner. It would be interesting to study the transition between these regimes, its impact
on capillary condensation and also understand its mesoscopic rounding when thermal fluctuations are included. Indeed, it is natural to think that
this is related to the commonly observed phenomena of contact angle hysteresis. The distinction between type I and type II condensation involving
pinned and unpinned menisci is also pertinent to other geometries, e.g., if we bring a vertical cylinder towards a macroscopic surface. The rounding
of the meniscus transition considered here arises due to the thermal fluctuations of the adsorbed wetting layers and occurs for even perfectly sharp
geometries. It would also be interesting to understand how surface roughness affects the edge contact angle and the meniscus depinning transition,
which may well connect with the phenomena of contact angle hysteresis. Including gravity may also introduce interesting new effects associated with
capillary emptying transitions \cite{dirk1, dirk2}. Finally, the equilibrium phase transitions considered here are also a pre-requisite for
understanding the dynamics of meniscus depinning which may be studied, for example, using dynamical DFT or simulation methods similar to those
described in \cite{trobo}.

\begin{acknowledgments}
\noindent This work was financially supported by the Czech Science Foundation, Project No. GA 20-14547S.
 \end{acknowledgments}

\end{document}